\documentclass[usenatbib,useAMS]{mn2e}

\usepackage{graphicx,subfigure}

\title[OH suppression with FBGs]{Suppression of the near-infrared OH night sky lines with fibre Bragg gratings  -- first results}

\author[Ellis et al.]{S. C. Ellis$^{1,2}$\thanks{E-mail: sellis@aao.gov.au},  J. Bland-Hawthorn$^{2}$,  J. Lawrence$^{1}$, A. J. Horton$^{1}$, C. Trinh$^{2}$, \newauthor  S. G. Leon-Saval$^{2}$,  K. Shortridge$^{1}$,  J. Bryant$^{2}$,  S.~Case$^{1}$, M. Colless$^{1}$, W.~Couch$^{3}$,  
\newauthor K.~Freeman$^{4}$, L.~Gers$^{1}$, K.~Glazebrook$^{3}$, R. Haynes$^{5}$, S. Lee$^{1}$, H.-G.~L\"{o}hmannsr\"{o}ben$^{6}$,
\newauthor   J. O.Byrne$^{2}$, S. Miziarski$^{1}$, M.~Roth$^{5}$, B.~Schmidt$^{4}$, C. G.~Tinney$^{7}$ and J. Zheng$^{1}$\\
$^{1}$Australian Astronomical Observatory, P.O. Box 296, Epping, NSW 1710, Australia\\
$^{2}$Sydney Institute for Astronomy, School of Physics, University of Sydney, NSW 2006, Australia\\
$^{3}$Centre for Astrophysics and Supercomputing, Swinburne University of Technology, PO Box 218, Hawthorn, VIC 3122, Australia\\
$^{4}$Research School of Astronomy and Astrophysics, Australian National University, Weston Creek, ACT 2611, Australia\\
$^{5}$innoFSPEC - Leibniz-Institut f\"{u}r Astrophysik Potsdam, An der Sternwarte 16, 14482 Potsdam, Germany\\
$^{6}$innoFSPEC - Institut f\"{u}r Chemie/Physikalische Chemie, Universit\'{a}t Potsdam, Karl-Liebknecht-Strasse 24-25, D-14476 Golm, Germany\\
$^{7}$School of Physics, University of New South Wales, Sydney 2052, Australia} 
\def\lsim{\mathrel{\hbox{\rlap{\hbox{\lower4pt\hbox{$\sim$}}}\hbox{$<$}}}}
\def\gsim{\mathrel{\hbox{\rlap{\hbox{\lower4pt\hbox{$\sim$}}}\hbox{$>$}}}}
\def\um{$\mu$m}
\def\bright{photons s$^{-1}$ m$^{-2}$ arcsec$^{-2}$ $\mu$m$^{-1}$}

\date{Accepted...... Received .....}

\begin{document}
\maketitle

\begin{abstract}
The background noise between 1 and 1.8 \um\ in ground-based instruments is dominated by atmospheric emission from hydroxyl molecules.  We have built and commissioned a new instrument, GNOSIS, which suppresses  103 OH doublets between $1.47$ -- $1.7$\um\ by a factor of $\approx 1000$ with a resolving power of $\approx 10,000$.  We present the first results from the commissioning of GNOSIS using the IRIS2 spectrograph at the Anglo-Australian Telescope.  We present measurements of sensitivity, background and throughput.  The combined throughput of the GNOSIS fore-optics, grating unit and relay optics is $\approx 36$ per cent, but this could be improved to $\approx 46$ per cent with a more optimal design.  We measure strong suppression of the OH lines, confirming that OH suppression  with fibre Bragg gratings will be a powerful technology for low resolution spectroscopy.  The integrated OH suppressed background between 1.5 and 1.7 $\mu$m is reduced by a factor of 9 compared to a control spectrum using the same system without suppression.  The potential of low resolution OH suppressed spectroscopy is illustrated with example observations of Seyfert galaxies and a low mass star.

The GNOSIS background is dominated by detector dark current below 1.67 $\mu$m and by thermal emission above 1.67 $\mu$m.  After subtracting these we detect an unidentified  residual  interline component of $\approx 860 \pm 210$ \bright, comparable to previous measurements.  This component is equally bright in the suppressed and control spectra.  We have investigated the possible source of the interline component, but were unable to discriminate between a possible instrumental artifact and  intrinsic atmospheric  emission.  Resolving the source of this emission is crucial for the design of fully optimised OH suppression  spectrographs.  The next generation OH suppression spectrograph will be focussed on resolving the source of the interline component, taking advantage of better optimisation for a fibre Bragg grating feed incorporating refinements of design based on our findings from GNOSIS.  We quantify the necessary improvements for an optimal OH suppressing fibre spectrograph design.

\end{abstract}


\section{Introduction}

Observations at near-infrared wavelengths are severely hindered by the night-sky background.  The night-sky surface brightness is $\approx 14.9$  AB mag arcsec$^{-2}$ in the H band, compared to $\approx 21.1$ AB mag arcsec$^{-2}$ in the V band.  This results in high Poisson noise in any observation.  Compounded with this, the night-sky brightness varies by factors of $\approx 10$ per cent on the timescale of minutes (\citealt{ram92}; \citealt{fre00}), with a further gradual dimming of about a factor of 2 throughout the night (\citealt{shi70}).  This temporal variability results in a high systematic noise when performing sky subtraction, which is not trivial to remove (e.g.\ \citealt{dav07}; \citealt{sha10}).

In the J and H bands the dominant sources of background are the Meinel bands of emission lines resulting from the rotational and vibrational de-excitation of OH molecules (\citealt{mei50}; \citealt{duf51}).  We have reviewed the near-infrared background, and in particular the characteristics of the OH emission line spectrum in an earlier paper (\citealt{ell08}).  We note that although they are very bright, the OH lines are intrinsically narrow (FWHM $\approx 3 \times 10^{-7}$\um), and between the OH lines the night-sky background should be very faint, possibly as low as $\approx 100$ \bright\ if the background is dominated by zodiacal scattered light, and is likely to be at least as faint as $\approx 600$ \bright\ as shown by $R=17,000$ spectroscopic observations made by \citet{mai93}.  
Thus, if the OH lines can be efficiently filtered from the night-sky spectrum, whilst maintaining good throughput between the lines, it should be possible to achieve a dark background in the NIR, allowing much deeper observations than possible hitherto.  

This paper presents the first results from an instrument designed to achieve exactly this efficient filtering of the OH skylines.  The filtering is achieved using fibre Bragg gratings (FBGs).    FBGs were originally developed for use in telecommunications, and required significant modification to be used for OH suppression  (see \citealt{bland11} for a comprehensive treatment on the physics of OH suppression with FBGs).   Firstly, it was  necessary to significantly increase the number of notches in each grating (typically only one notch is used in devices used in telecommunications) and the wavelength range of the  devices which were available at the time, a breakthrough which was enabled with the design of aperiodic FBGs (\citealt{bland04}).  Secondly it was required to develop a multi-mode to single-mode fibre converter (\citealt{leo05,noo09}), since at the plate-scale of typical telescopes the narrow core of a single mode fibre has too small a field-of-view to collect an adequate amount of light from the seeing disc.    Subsequent refinement (\citealt{bland08,bland11}) has resulted in the latest FBGs being able to suppress 103 notches over a wavelength range of 230~nm, which is achieved using two devices in series.  

\cite{ell08} presented the potential benefit for astronomy of OH suppression with FBGs and described the expected performance.  \citet{ell08} point out that a key benefit of OH suppression with FBGs compared to other previously suggested methods of OH suppression (e.g.\ high dispersion masking, \citealt{mai00}) is that the OH light is removed before it enters the spectrograph and in a manner dependent only on wavelength.  Hence the interline continuum is not contaminated with scattered OH light within the spectrograph which can otherwise dominate other sources of interline continuum.  The testing of this prediction was a central goal of GNOSIS commissioning.

The first on-sky tests of OH suppression with FBGs were performed at the Anglo-Australian Telescope in December 2008.  This experiment consisted of two multimode fibres pointed directly at the sky through hole in the AAT dome wall.  Both fibres fed a $1 \times 7$ photonic lantern (see \S~\ref{sec:inst}), one of which had FBGs inserted, and the other did not, to serve as a control.  The results of these tests are described in \citet{bland09,bland11}, and demonstrated the clean suppression of the OH lines.  However, since each fibre accepted light from a $\approx 12$ degree patch of sky these tests could not perform observations of individual sources, nor could they measure the interline continuum since every
observation included light from many stars and other sources.

We have designed, built and commissioned the first instrument to use FBGs for OH suppression.  GNOSIS is an OH suppression unit designed to feed the IRIS2 infrared imaging spectrograph at the AAT (\citealt{tin04}).  The name is an acronym for Gemini Near-infrared OH Suppression IFU System, revealing the intention to install a similar system on the GNIRS spectrograph at Gemini North (\citealt{eli98,eli06,eli06b}) at a future date.  
A summary description of the instrument is given in section~\ref{sec:inst} below.  In this paper we concentrate on the performance and results of the OH suppression.  We describe the observations in section~\ref{sec:obs} and we present the  results   in section~\ref{sec:results} giving details of throughput, sensitivity and background.  The near-infrared background will be examined in detail in a future paper (Trinh et al., in prep.), examining the various components of the background and their dependence on moonlight, ecliptic latitude, airmass, Galactic latitude etc.; a brief description of the main components will be given here.  In section~\ref{sec:sciobs} we present  observations of two Seyfert galaxies and an L/T dwarf illustrative of OH suppression.  In section~\ref{sec:disc} we discuss our results in the context of the benefit of OH suppression to NIR spectroscopy and in the context of lessons for future OH suppressed instruments.

\section{GNOSIS}
\label{sec:inst}

\begin{figure*}
\centering \includegraphics[scale=0.6]{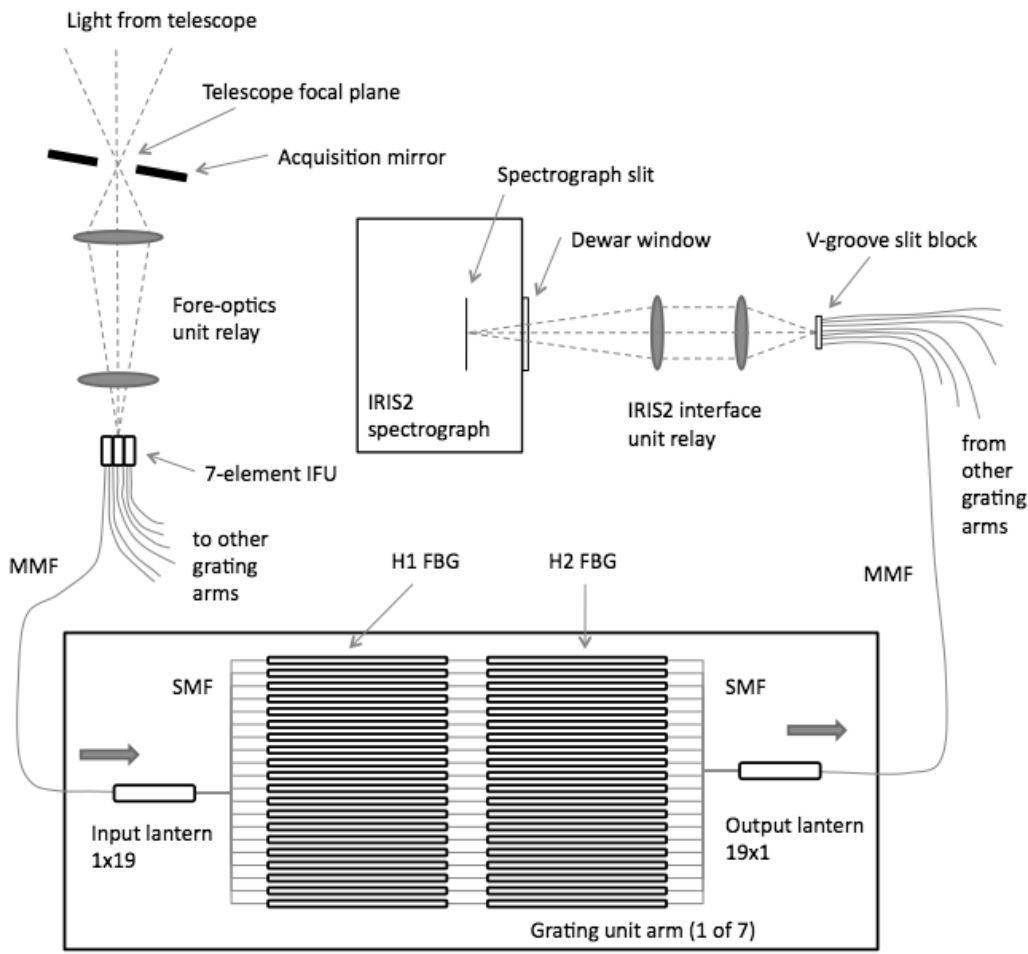}
\caption{A schematic diagram of the components of the GNOSIS instrument, as described in section~\ref{sec:inst}.}
\label{fig:gnosis}
\end{figure*}

The optical light path for GNOSIS is illustrated in Figure~\ref{fig:gnosis}.  A 7 element lenslet array accepts light from the f/8 Cassegrain focus and feeds this to seven 50\um\ core fibres.  The individual lenslets subtend a width (from face to face) of 0.4 arcsec on the sky.  Each of the 50\um\ core fibres is capable of carrying $\approx 79$ modes, but light is injected at less than the full numerical aperture of the fibres (i.e. a slower beam) such that only  $\approx 19$ modes are carried.  These fibres are spliced to a photonic lantern (\citealt{leo05,noo09}), which converts a multimode fibre (MMF) into a parallel array of single mode fibres (SMFs) via a fibre taper; in the case of GNOSIS into 19 single mode fibres since there are 19 modes per fibre.  These single mode fibres are each spliced to two FBGs in series which suppress the 206 brightest OH lines (i.e.\ 103 closely spaced $\Lambda$ doublets) between $1.47$ and 1.7 \um.  These FBGs are then spliced to a reverse photonic lantern converting each of the $7 \times 19$ SMFs back into 7 MMFs.  These are connected to a 12m fibre run which leads to IRIS2, located on the dome floor beneath the telescope.  The ends of the fibres form a pseudo-slit, the output of which is reimaged via two lenses into IRIS2.  A custom slit mask within IRIS2 blocks extraneous light.  IRIS2 has a Rockwell Hawaii-1 detector with $1024 \times 1024$ pixels, a dark current of 0.015 e$^{-}$ s$^{-1}$ and an effective read noise of $\approx 8$ e$^{-}$ when using an up-the-ramp non-destructive read-out mode.  The point spread function of GNOSIS with IRIS2 was measured to be 2.0 pixels FWHM, which gives a spectral resolving power of $\lambda/\Delta\lambda \approx 2350$.  The mean dispersion was measured to be $\approx 3.5$ \AA\ pixel$^{-1}$.
A full description of the instrument may be found in a forthcoming paper (Trinh et al.\ in prep.).  

The properties of the GNOSIS system are given in Table~\ref{tab:gnosis}.  These include properties of GNOSIS {\it per se} and also those properties of IRIS2 on which  GNOSIS observations depend.  The observed wavelength range is defined by the spectrograph and the blocking filter, and is larger than the OH suppressed range which is that of the FBGs.

\begin{table}
\caption{Properties of the GNOSIS system.}
\label{tab:gnosis}
\begin{tabular}{ll}
Total field of view & 0.97 arcsec$^2$ \\
Field of view per IFU element & 0.14 arcsec$^{2}$\\
Spectral resolving power & 2350 \\
PSF FWHM & 2 pixels\\
Dispersion & $\approx 3.5$ \AA\ pix$^{-1}$\\
Observed wavelength range & 1.49 -- 1.77\um \\
OH suppressed range & 1.47 -- 1.7\um \\
Read noise (MRM)& $\approx 8$ e$^{-}$ \\
Dark current & $\approx 0.015$ e$^{-}$ s$^{-1}$
\end{tabular}
\end{table}

\section{Observations}
\label{sec:obs}

GNOSIS was commissioned on five separate observing runs in March, May, July, September and November of 2011.  In this paper we present only observations from the September and November commissioning runs, which incorporate the improvements made to the instrument in the previous commissioning runs.  The observations presented are listed in Table~\ref{tab:obs}.  We present four different experiments: (i) measurement of the instrument throughput from standard star observations, (ii) measurement of the night sky background and (iii) measurement of the instrument sensitivity by observations of a low-surface brightness galaxy, and (iv) observations of Seyfert galaxies illustrating the benefits of OH suppression.

All observations were made using an up-the-ramp read mode to minimise the effect of detector read noise; we have recorded the period and number of reads in Table~\ref{tab:obs}; the final exposure time for an individual frame is $({\rm reads} - 1) \times {\rm period}$.    All observations of objects included an equal amount of time spent on sky, and followed an object-sky-sky-object pattern.  We give the total number of object frames as ``Nods'' in Table~\ref{tab:obs}.  For observations on 1st -- 4th September we disconnected a MMF from the grating unit, to provide a control fibre with no OH suppression, as indicated in the  `Control' column of Table~\ref{tab:obs}.

\begin{table*}
\caption{GNOSIS observations}
\label{tab:obs}
\begin{tabular}{llllllll}
Object & Date & Period (s) & Reads & Nods & Total exposure (s) & Control & Comments \\ \hline
Sky & 1st Sep 2011 & 30 & 61 & 6 & 10800 & Y &  \\
Sky & 2nd Sep 2011 & 30 & 61 & 3 & 5400 & Y &  \\
HIP 104664 & 3rd Sep 2011 & 5 & 13 & 2 & 120 & Y & A0V standard star, $H=8.6$ mag\\
HIP 109476 & 5th Sep 2011 & 5 & 13 & 2 & 120 & N & A0V standard star, $H=7.9$ mag\\
Sky & 3rd Sep 2011 & 30 & 61 & 4 & 7200 & Y &  \\
Sky & 4th Sep 2011 & 30 & 61 & 2 & 1800 & Y &  \\
HIZOA J0836-43 & 26th Nov 2011 & 15 & 61 & 2 & 1800 & Y & LSBG$^{1}$, $H=17.3$ mag arcsec$^{-2}$\\
NGC 7674 & 4th Sep 2011 & 15 & 61 & 2& 1800 & Y & Seyfert galaxy\\
NGC 7714 & 4th Sep 2011 & 15 & 61 & 2 &1800 & Y & Seyfert galaxy\\
2MASS 0257-3105 & 27th Nov 2011 & 15 & 61 & 1 & 900 & N & Candidate L7 dwarf \\
\multicolumn{8}{l}{{\scriptsize1. Low surface brightness galaxy.}}
\end{tabular}
\end{table*}

\subsection{Data reduction}
\label{sec:datred}

We reduced the data using bespoke code written as a {\sc Mathematica} notebook, except for the first step of collapsing the data cubes (see \S~\ref{sec:cube}), which was performed using a C program.  Since the night sky background is considerably reduced using OH suppression the issue of correctly accounting for the effects of detector noise becomes very important, and we have paid careful attention to this in our data reduction.  Our procedure was as follows.

\subsubsection{Collapsing the data cubes}
\label{sec:cube}

\begin{figure}
\begin{center}
\subfigure[MRM observations]{
\includegraphics[scale=0.6]{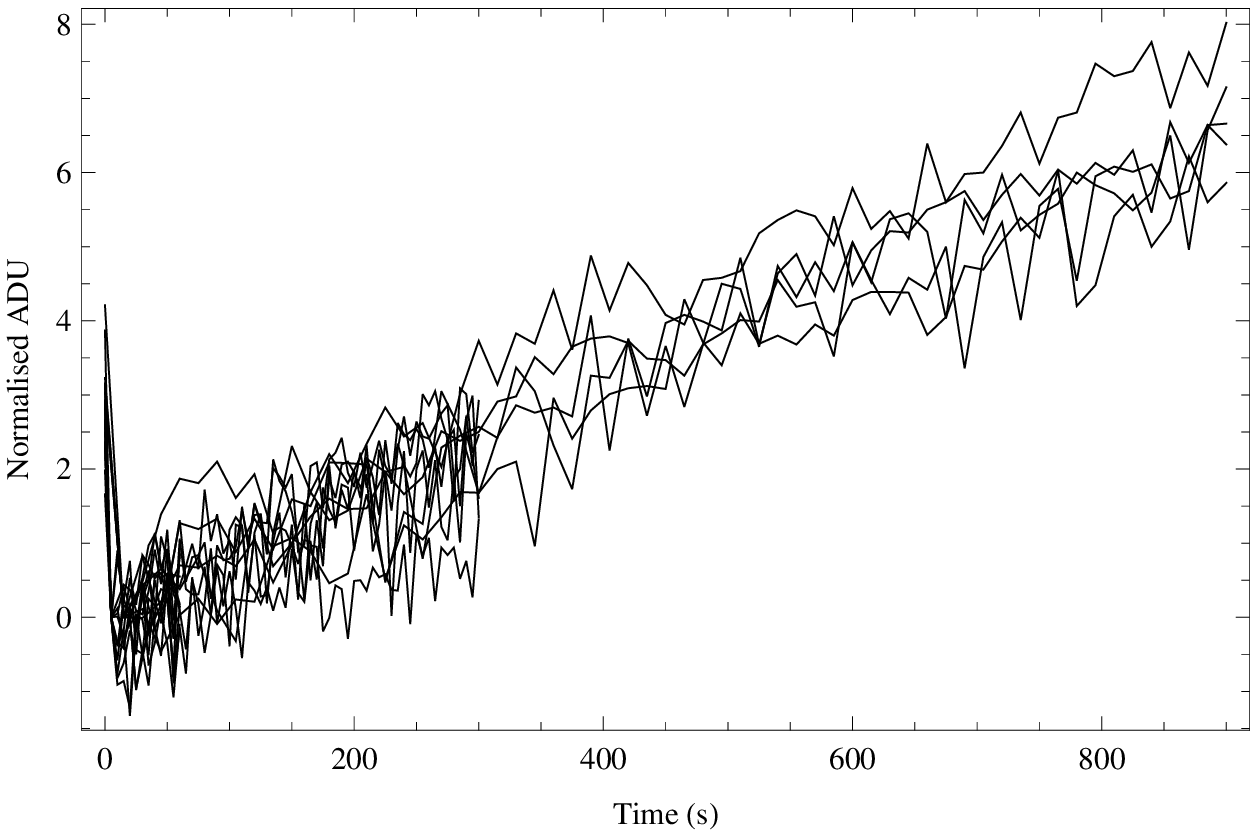}
}
\subfigure[DRM then MRM observations]{
\includegraphics[scale=0.6]{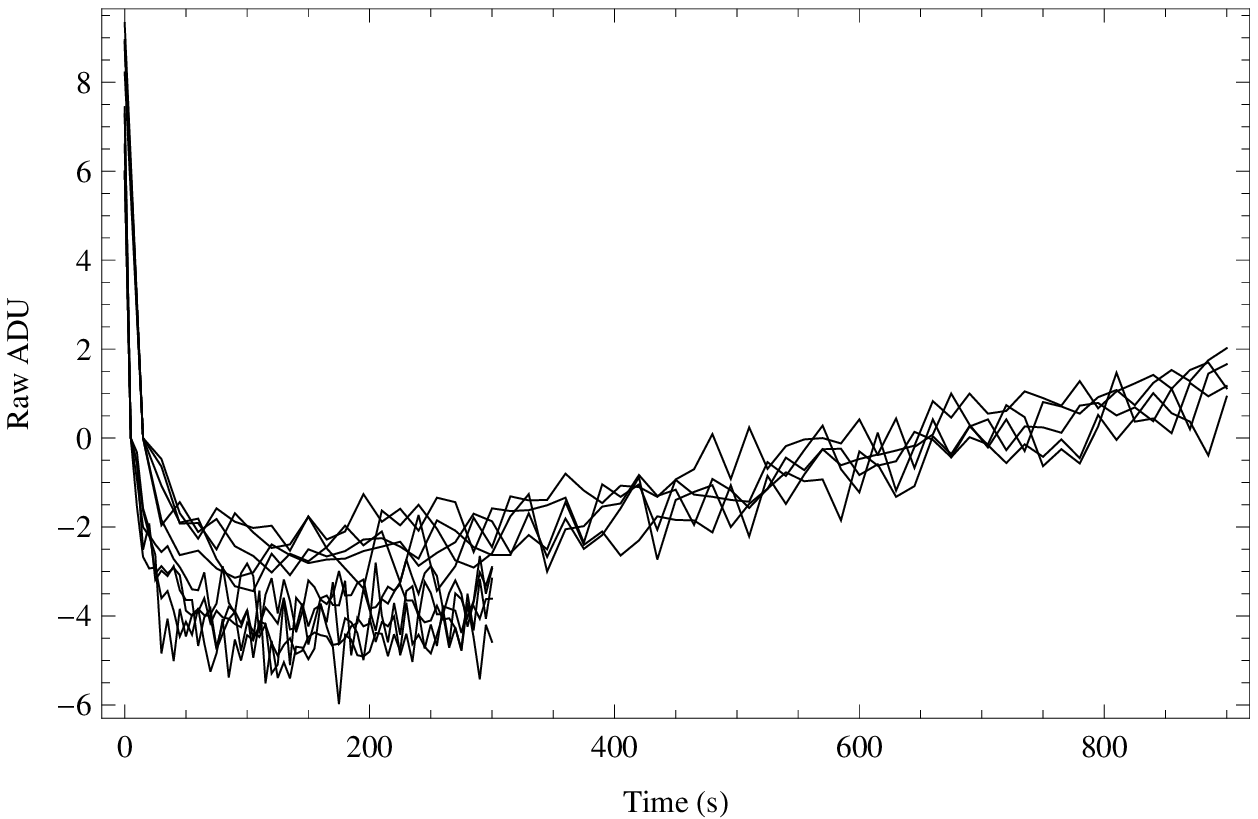}
}
\caption{The dark current in raw ADU for the IRIS2 detector.  The data have been normalised to zero for the second read.  The top plot shows data if the detector is always read in MRM mode.  The bottom plot shows data if the MRM frame is preceded by a DRM frame.}
\label{fig:dark}
\end{center}
\end{figure}

For each frame the detector was used in a multiple-read mode (MRM) using an up-the-ramp sampling, with intervals of either 5, 15 or 30s and either 13 or 61 reads, see Table~\ref{tab:obs}.  This means that the detector is re-set, then read non-destructively at a fixed time interval for a specified number of times.  This enables non-linearity in the detector response to be removed.  Upon investigation we found significant non-linearity in the first few reads.  For dark frames with the blank filters in, if one always observes  in MRM then the first read after re-set is higher than the second read, and thereafter the counts are linear until about $20,000$ ADU (see Figure~\ref{fig:dark}a).  If the MRM exposure follows a double read mode observation (DRM; i.e. only two reads immediately prior to and after the exposure), then the first $\sim 5$ reads show a decline in the bias levels (see Figure~\ref{fig:dark}b).  This effect has been seen by one of us (AJH)  in other Hawaii-1 PACE detectors.  The largest observed drop in bias levels in the first few reads is $\approx 15$ ADU, which is insignificant under typical non-OH suppressed observations since the brightness of the night sky swamps such small  variations.  Thus for normal NIR observations a linear least-squares fit to the reads suffices to recover the incident count rate.  However, in the case of OH suppressed observations this is no longer so, since the incident count rate can be very low for pixels between the OH lines.  Therefore we have simply dropped the first 5 reads from our data cubes, before performing a linear least squares fit to the remaining data.  The slope of this fit is then multiplied by the original number of reads minus one to recover the true incident count rate.

\subsubsection{Detector linearity}

The detector non-linearity due to the filling of the pixel wells is mitigated by keeping the counts per pixel below 20,000 ADU.  Thus in collapsing the data cubes as described above, we use a linear least squares fit.  Thereafter we correct for any non-linearity following the procedure given on the IRIS2 web pages (http://www.aao.gov.au/AAO/iris2/iris2\_linearity.html).

\subsubsection{Background subtraction}
\label{sec:backsub}

The GNOSIS background consists of dark current, instrument thermal background, telescope thermal background, and the night sky background (including OH emission, thermal emission, zodiacal scattered light, moonlight etc.).  For our object observations we nod the telescope between object and sky, and thus all the background can be subtracted simultaneously using the sky observations.  Pairs of observations are first subtracted, then averaged using the median if there is more than two.

For our sky observations this is not possible.  In this case we subtract the dark current and the thermal background separately.  The dark current is subtracted using the median of several dark frames of the same read mode and exposure time as the observations.

We have measured the thermal component from ``cold frame'' observations.  A cold frame is an observation for which the GNOSIS fore-optics were removed from the Cassegrain focus and pointed directly into a dewar filled with liquid nitrogen.  Thus the cold frame contains the GNOSIS fore-optics and IRIS2 thermal backgrounds and the detector dark current alone.  The telescope thermal background should be much lower than the instrument background since the emissivity from the mirror should be less than $3$ per cent, and the emissivity from the Cassegrain hole is from a much smaller solid angle.  

\begin{figure}
\centering \includegraphics[scale=0.6]{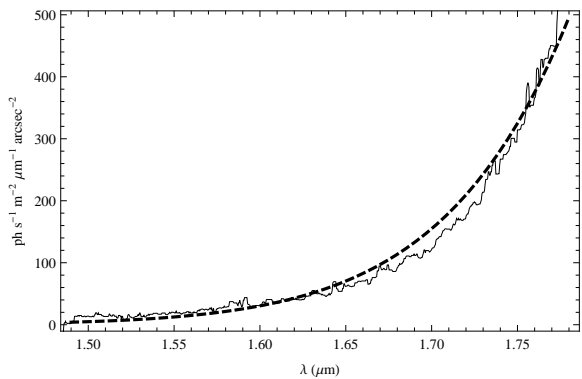}
\caption{The smoothed GNOSIS thermal background (continuous line) normalised to a $T=282$ K black body, i.e.\ the ambient dome temperature at the time of observation (dashed line).}
\label{fig:cold}
\end{figure}

The average cold frames are shown in Figure~\ref{fig:cold}.  The counts have been calibrated to  reproduce a $T=282$ K, 100 per cent emissivity black body, i.e.\ the expected emission from the slit block etc.\ at the output end of GNOSIS.  This calibration provides an independent check of the efficiency of IRIS2 as discussed in section~\ref{sec:through}.

\begin{figure}
\centering \includegraphics[scale=0.6]{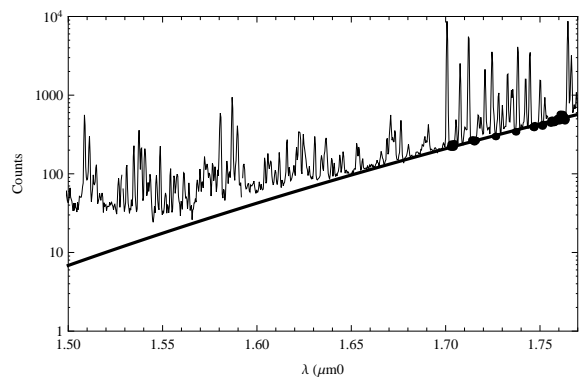}
\caption{An example fit (thick, black line) to the GNOSIS thermal background which is taken to be given by the points located between the OH lines in the night sky spectrum (thin line).}
\label{fig:bbody}
\end{figure}

Obtaining cold frames is a significant overhead on observing time since the fore-optics must be removed and then re-fitted and re-aligned during the night when the optics are at the same temperature as for the sky frames.   Thus in practice we fit a black-body spectrum to our observations by choosing regions of the continuum between the OH lines with $\lambda > 1.7$\um.  
The best fitting model is subtracted from observations.  
We note that thermal emission is only approximately black-body (see Figure.~\ref{fig:cold}).  However, a more general function may inadvertently fit and subtract  intrinsic features of the sky emission.  Since we do not know the exact function of the thermal emission we revert to an approximate, but physically motivated black-body fit. 
An example black-body fit is shown in Figure~\ref{fig:bbody}.  This is actually done after the residual background correction step detailed in ~\S~\ref{sec:residback} below.  

\subsubsection{Spectral extraction}

Spectroscopic observations of the illuminated dome flat field screen were taken to provide a trace of the spectra across the detector.  The centre of each of the 7 spectra is calculated at
each spectral pixel.  This is done by fitting the sum of seven Gaussians to the pixel counts in the spatial direction at each spectral pixel.   The Gaussians for each fibre are assumed to have identical $\sigma$ and to be equally spaced.  The fit yields the centre of each Gaussian, and this is stored in a look-up table, along with the $\sigma$ of the Gaussian fit.  The separation between the spectra is 11 pixels, which is $\approx 5.5$ times the FWHM of the PSF, and therefore cross-talk between the spectra is negligible.

Object spectra are then extracted  following the ``Gaussian summation extraction by least squares'' method of \cite{sha10b}, i.e.\ the count for each spectral pixel, $c$ is given by,
\begin{equation}
c=\frac{\sum_{i}\frac{D_{i} \phi_{i}}{\sigma_{i}^{2}}}{\sum_{i}\frac{\phi_{i}^{2}}{\sigma_{i}^{2}}}
\end{equation}
where the sum ranges over each pixel,  $i$,  in the spatial direction between $\pm \frac{1}{2}$ the pitch between spectra, $D_{i}$ is the count at each pixel, $\sigma_{i}^{2}$ is the variance on the count, and $\phi_{i}$ is the value of the normalised Gaussian profile at that pixel.  This method minimises the contribution due to noise in the extraction.

\subsubsection{Fibre-to-fibre variations}

We next correct for variations in throughput between fibres.  The fibre-to-fibre variation is measured from dispersed images of the illuminated dome flat field screen which 
are extracted in the same manner as the object spectra.
The normalised fibre-to-fibre variation is the ratio of the total flux in each spectrum to the average of the total flux of all seven spectra.  The extracted spectra are divided by the normalised  variation.

\subsubsection{Spectral combination}

The seven extracted spectra are combined by taking the sum of each spectral pixel.  If a control fibre was used, this is excluded from the combination.  This step neglects to take into account differences in the wavelength solution of each spectrum, which are  less than 1.5 pixels over most of the detector.  We have checked that not correcting for these differences does not introduce spurious signals by repeating the reduction of the sky spectrum for a single fibre only. We find that the single extracted spectrum is consistent with that of the summed spectrum.

\subsubsection{Wavelength calibration}

The wavelength calibration is achieved using Xe arc lamp spectra.  These spectra are extracted and combined as described above.  The Xe lines are automatically detected and a cubic polynomial is fit to give the wavelength as a function of spectral pixel number.  This solution is then applied to the object spectra.  The wavelength calibration is accurate to $\approx 2$ \AA, and is limited by the slight shift in the dispersion solution for each spectrum.

\subsubsection{Residual background subtraction}
\label{sec:residback}

Hawaii-1 detectors are known to suffer from significant inter-quadrant cross-talk (\citealt{tin03}\footnote{See also {\scriptsize \\http://www.eso.org/$\sim$gfinger/hawaii\_1Kx1K/crosstalk\_rock/crosstalk.pdf}}); if a bright object appears at pixel $\left\{ x,y \right\}$, then there will also be a faint glow at pixel $\left\{ x,{\rm mod}(y+512,1024)\right\}$ (where ${\rm mod}$ is the modulo operation), and along the entire row $x$.  The cross-talk between quadrants does not affect GNOSIS observations, since all seven spectra are located on the lower two quadrants, but the cross talk along rows does.  

The cross-talk along a row can be measured in the region of the spectrum at $\lambda < 1.45$\um, since this region ought to be completely dark due to the spectroscopic blocking filter.  This measurement may also include other residual background from systematic errors in the previous subtraction, due to changing background levels.  We measure the mean count rate at $\lambda < 1.45$\um\ and subtract it from the spectrum.

\subsubsection{Instrument response and telluric correction}

The efficiency of the GNOSIS system, including the atmosphere, the telescope and IRIS2, is measured from observations of A0V stars.  These observations are reduced as described in the previous steps.  We then take a model spectrum of Vega (\citealt{cas94}) and divide the observations by this to give a relative throughput of the system as a function of wavelength, including telluric features.  Note that this will also include the averaged pixel-to-pixel variation in our extracted spectra, and hence includes the detector flat-field response.  The relative throughput is normalised between 1.5 and 1.69 \um\ to give the instrument response function.  The object spectra are then divided by the instrument response function.

\subsubsection{Flux calibration}
\label{sec:fluxcal}

The observations of the telluric standard stars described in the previous step can also be used to estimate the flux calibration.   Since the magnitude of the A0V stars is already known, we can scale the model Vega spectrum to the appropriate brightness before dividing the observed spectrum to yield a flux calibrated correction.  This step is not very accurate since scaling the Vega spectrum relies on knowing the aperture losses of the IFU at the time of observation, which in turn depends upon the seeing and acquisition, which are not readily measurable.   The flux calibration is discussed further in \S~\ref{sec:starthroughput}.

\begin{figure}
\centering \includegraphics[scale=0.6]{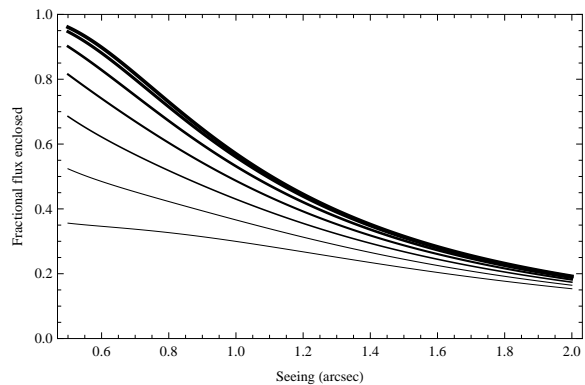}
\caption{The fractional flux enclosed in the IFU as a function of the seeing.  The different curves are for a point source offset from the centre of the IFU by 0, 0.1, 0.2, 0.3, 0.4, 0.5 and 0.6 arcsec going from the thickest to the thinnest lines.}
\label{fig:aploss}
\end{figure}

Due to the difficulty of measuring the seeing and the centring of our objects in the IFU we have computed the aperture losses for a seven element hexagonal array, assuming the seeing profile is Gaussian.  The results are shown in Figure~\ref{fig:aploss} as a function of seeing for several values of mis-alignment.  Assuming that we can accurately align to within one element of the IFU, i.e.\ the offset is $\le 0.2$ arcsec, and the seeing is 1.2 arcsec (extrapolated from measurements taken with the acquisition camera in the I band during the observations), then the typical aperture loss for a point source is $\le 0.58$.

\section{Results}
\label{sec:results}

We now describe the results of the first three experiments, viz.\ the instrument throughput, the night sky background and the instrument sensitivity.

\subsection{Instrument throughput}
\label{sec:through}

\subsubsection{Laboratory measurements}

The throughput of GNOSIS was measured in the laboratory using a super-continuum source and a near-infrared camera (see Trinh et al.\, in preparation for details).  The total end-to-end throughput of GNOSIS is $\approx 36$ per cent and the throughputs of the individual components are given in Table~\ref{tab:loss} with the estimated errors.

The throughput of the AAT is approximately $0.94^{2}$ for the two reflections off the aluminium coated primary and secondary mirrors, which is estimated from a $0.98$ reflectivity of bare Al, combined with an extra $0.96$ throughput from the accumulation of dirt on the mirrors etc.  This extra loss is compatible with that measured in the visible on the AAT.

We have estimated the throughput of IRIS2 in two ways.  First we used the IRIS2 imaging exposure calculator to estimate the efficiency of an imaging observation, which for H is $\approx 29$ per cent.  Combined with the throughput of the grism ($\approx 40$ per cent) this gives a total efficiency of $\approx 12$ per cent.  Secondly we used the observations of the ``cold frames'' (see \S~\ref{sec:backsub}) compared to a theoretical blackbody spectrum taking into account the area and solid angle of the IRIS2 slit-mask holes and assuming $T=282$ K as recorded, and 100 per cent emissivity, since most of the emission will come from the slit block and other mechanical parts surrounding the fibres due to the oversized slit mask holes.    This also yielded an efficiency of 12 per cent.  

The end-to-end system throughput is therefore expected to be, $\approx 3.6$ per cent, or $\approx 1.5$ per cent for a point source including typical aperture losses.  We emphasise that this low throughput is not an intrinsic property of OH suppression systems, but is rather a consequence of retro-fitting an OH suppression unit to an existing spectrograph, which already has only 12 per cent throughput.  The throughput of  GNOSIS itself is 36 per cent and is limited mainly by the photonic lanterns.  This will be discussed further in \S~\ref{sec:disc}.

\subsubsection{Standard star measurements}
\label{sec:starthroughput}

\begin{figure}
\begin{center}
\subfigure[HIP 104664]{
\includegraphics[scale=0.6]{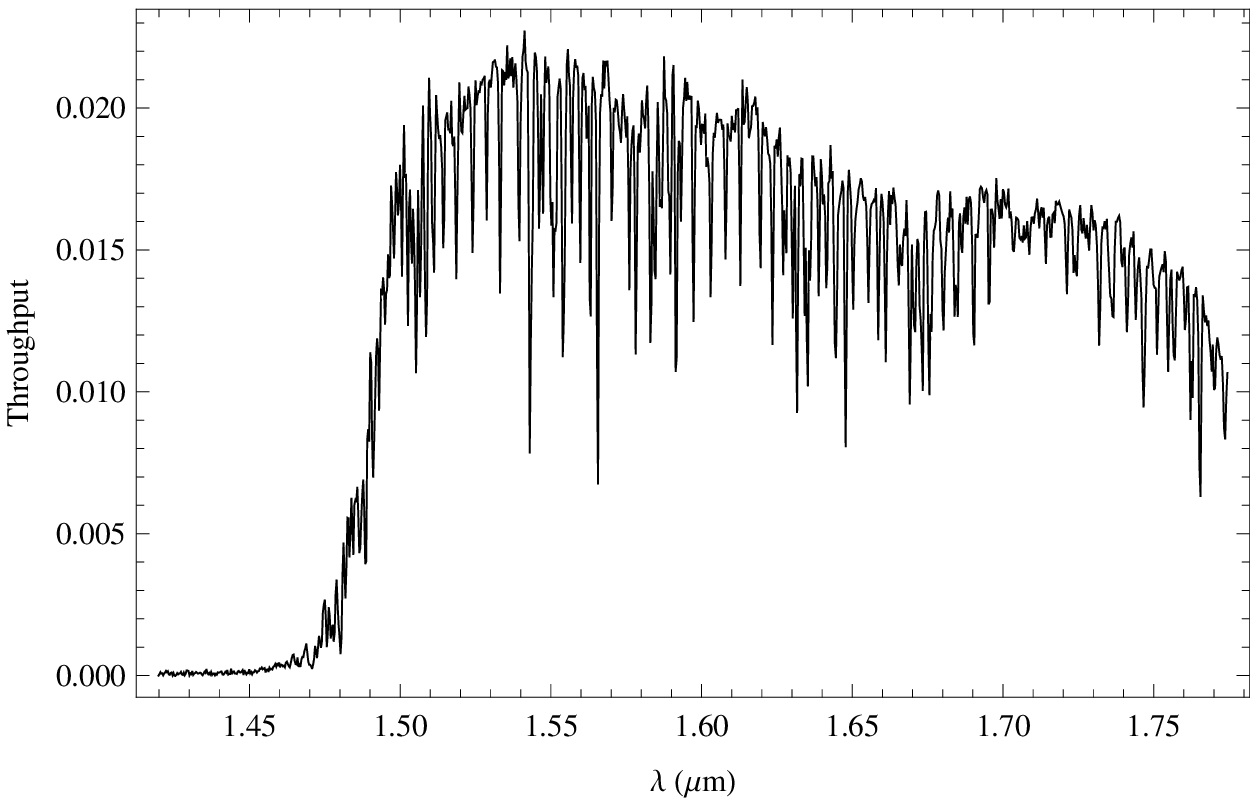}
}
\subfigure[HIP 109476]{
\includegraphics[scale=0.6]{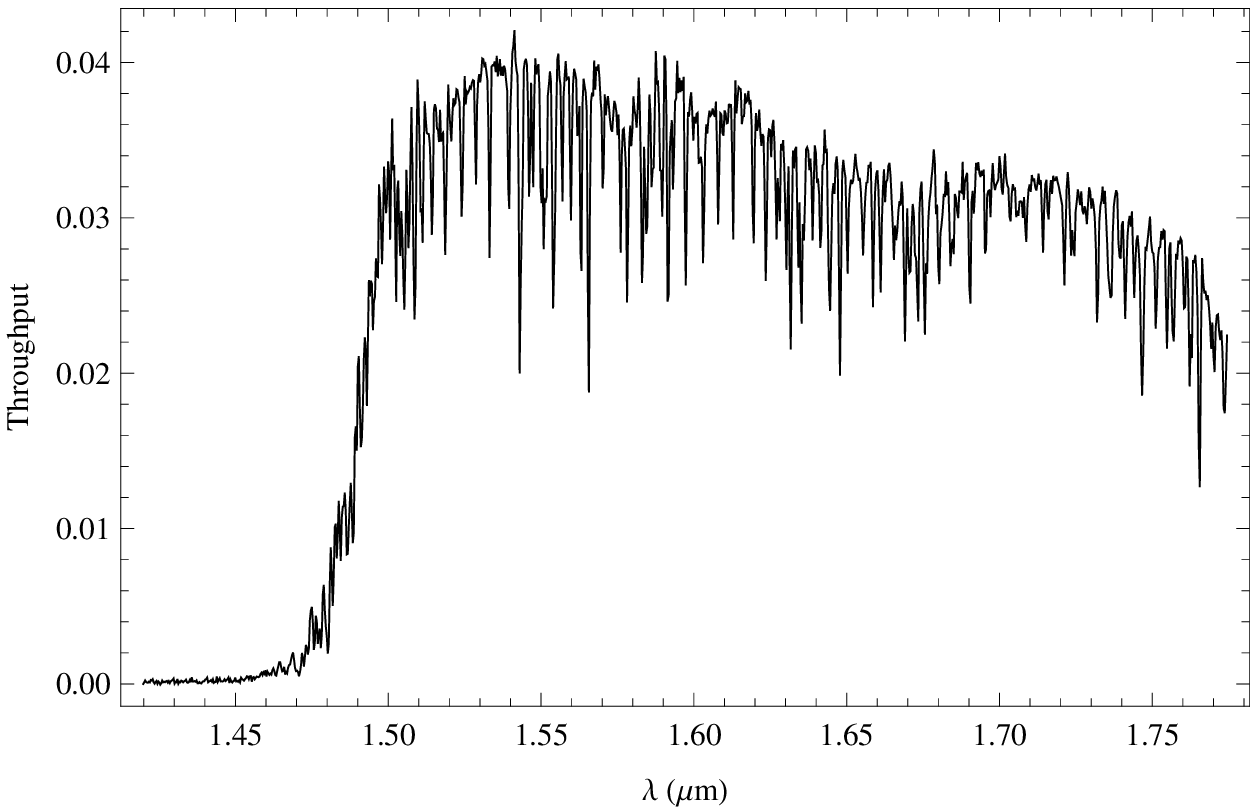}
}
\caption{The end-to-end throughput, including aperture losses and estimated errors, of the GNOSIS+telescope+IRIS2 system as measured from observations of A0V stars.  The small scale variation is real and is due to the FBGs (convolved with the spectrograph PSF).}
\label{fig:starthrough}
\end{center}
\end{figure}

We have also measured the end-to-end throughput as a function of wavelength from our observations of A0V stars.  Examples are shown in Figure~\ref{fig:starthrough}.  The notches in the FBGs are visible as the significant narrow dips in the throughput.  Note that in this graph the notches have been convolved with the spectrograph point spread function, so they appear shallower than the actual suppression depth.  The measurement for HIP 104664  does not include the control fibre.
The measured values are very sensitive to seeing and aperture losses as shown in Figure~\ref{fig:aploss}, but are within the errors of the anticipated throughput estimate in Table~\ref{tab:loss}.  In the case of HIP~109476 the measured throughput would require very low aperture losses due to exceptional seeing, suggesting some of the values in the table may be pessimistic.  However, we assume the conservative values in Table~\ref{tab:loss} for the rest of the paper.

\begin{table}
\begin{center}
\caption{Individual and cumulative throughput}
\label{tab:loss}
\begin{tabular}{lll}
Element & Throughput & Cumulative throughput \\ \hline
Primary mirror & $0.94 \pm 0.04$ & $0.94 \pm 0.04$ \\
Secondary mirror & $0.94 \pm 0.04$ & $0.88 \pm 0.05$\\
{\bf GNOSIS:} &  &  \\
\hspace{5mm} Fore-optics & $0.865 \pm 0.05$ & $0.76 \pm 0.06$\\
\hspace{5mm} IFU & $0.83 \pm 0.05$ & $0.63 \pm 0.07$\\
\hspace{5mm} Grating unit & $0.576 \pm 0.05$ & $0.37 \pm 0.05$ \\
\hspace{5mm} Relay-optics & $0.862 \pm 0.05$ & $0.31 \pm 0.05$\\
\hspace{5mm} \emph{Total} & \emph{0.356} $\pm$ \emph{0.05 } & \\
Relay alignment & $0.95 \pm 0.05$ &   $0.30 \pm 0.05$\\
IRIS2 & $0.12 \pm 0.10$ &  $0.036 \pm 0.03$\\
Aperture losses & $0.42 \pm 0.20$ & $0.015 \pm 0.015$ 
\end{tabular}
\end{center}
\end{table}

\subsection{Night sky background}

\begin{figure*}
\begin{center}
\subfigure[]{
\includegraphics[scale=1.2]{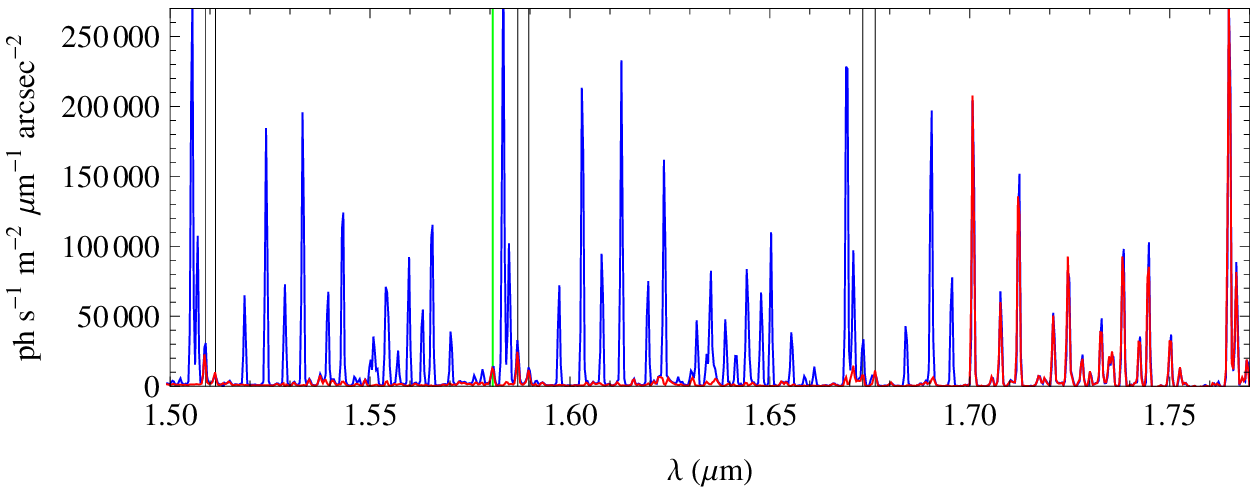}
}
\subfigure[]{
\includegraphics[scale=1.2]{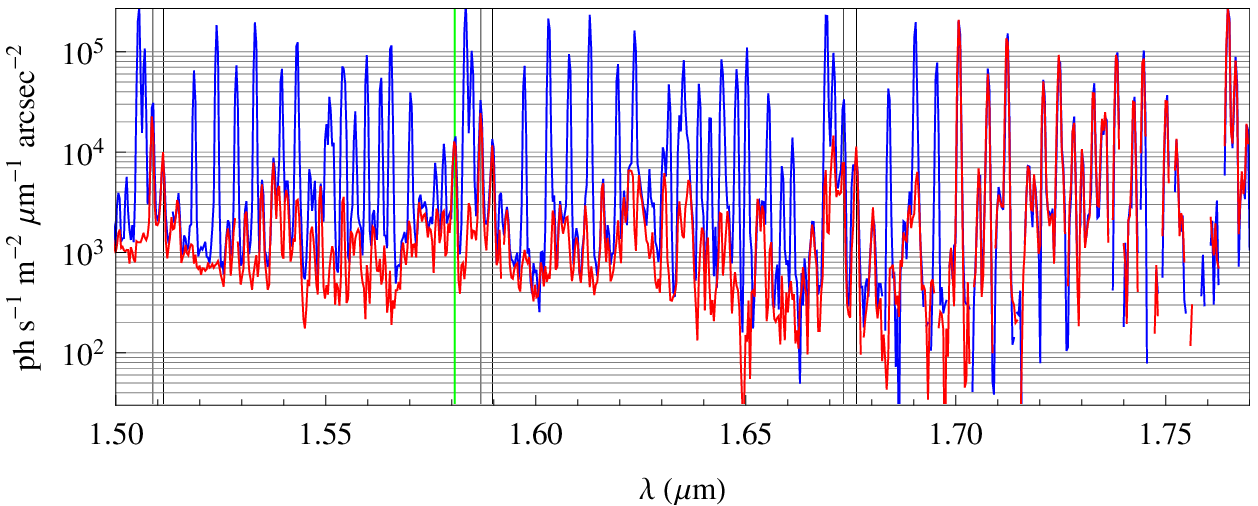}
}
\caption{The night sky spectra for the OH suppressed observations (red), and the control fibre without OH suppression (blue).  The two plots are of the same data, the bottom one has been logged to make the interline regions visible.  The pairs of vertical black lines show the OH lines from the Q1(3.5) and Q1(4.5) rotational lines in the 3 -- 1, 4--2 and 5--3 vibrational transitions (from left to right).  The green line shows the O$_{2}$ a-X v=0 -- 1 vibrational line.}
\label{fig:sky}
\end{center}
\end{figure*}

Observations of blank sky were made as targeted observations and as ``sky frames'' during observations of objects.   Figure~\ref{fig:sky} shows the reduced sky spectra for 7.5 hours of observations on the 1st -- 4th  of September.  These data were all  $>80$ deg from the moon, $>50$ deg south of the Galactic plane, $>15$ deg off the ecliptic plane and at an airmass  $<1.5$.  Therefore the night sky spectra should not be significantly contaminated by moonlight, starlight, or zodiacal scattered light.

\subsubsection{OH suppression}
\label{sec:ohsupp}

The OH lines between 1.5 -- 1.7 \um\ are strongly suppressed.  We have measured the suppression factor for 57 doublets; for the remaining 46 doublets we were unable to obtain a good fit to the unsuppressed line due to either line blending, intrinsic faintness or due to the filter cut-off (at $< 1.5\ \mu$m).  In most cases the measured strength of the suppressed line is poorly constrained because the flux is so low.  Nevertheless, of the 57 doublets 78 per cent meet or exceed the target specifications.

Some quite bright lines are not suppressed by the FBGs.  These are marked with black and green vertical lines in Figure~\ref{fig:sky}.  The green line is an O$_{2}$ emission line resulting from the a-X v=0 -- 1 vibrational transition (Jeremy Bailey, private communication).  This line is very bright in the day time and fades rapidly throughout twilight, but there is faint persistent emission throughout the night.  It was not included in our FBG designs.

The lines marked in black are OH lines from the Q1(3.5) and Q1(4.5) rotational lines in the 3--1, 4--2 and 5--3 vibrational transitions.  The FBGs have notches of $\approx 20$ dB for all these lines, printed at the wavelengths given by \citet{rous00}.  However, we measure a mean suppression factor of only $\approx 2.5$ dB.   
There may be an issue with the printed notch wavelengths of these particular transitions due to a larger spacing between the individual $\Lambda$ doublets than for other transitions (Pierre Rousselot, private communication).  We show the wavelengths and gap size of the transitions in question as modelled by \citet{rous00} and as measured by \citet{abr94} in Table~\ref{tab:qlines}.  The mean wavelengths of the doublets are identical in each case, but the measured gap size is considerably larger; this is not a problem for the Q1(0.5), Q1(1.5) and Q1(2.5) transitions for which the $\Lambda$ doublets are much more closely spaced ($< 100$ pm).   However for the Q(3.5) and Q(4.5) transitions the doublet spacings derived from the \citet{abr94} measurements are larger than the typical GNOSIS FBG notch widths, i.e.\ the individual lines fall either side of the notch.  These lines account for 37 per cent of the doublets which did not meet the target suppression depth.

\begin{table*}
\caption{The model (\citealt{rous00}) and measured wavelengths (\citealt{abr94}) for the 3--1, 4--2, 5--3, Q1(3.5) and Q1(4.5) transistions in $\mu$m.}
\label{tab:qlines}
\begin{tabular}{ccccccccc}
&\multicolumn{4}{c}{Rousselot} & \multicolumn{4}{c}{Abrams} \\  \hline
& e & f& mean & $\Delta\lambda$ & e & f& mean & $\Delta\lambda$ \\ \hline \hline
\multicolumn{9}{c}{3--1}\\ \hline
Q1(3.5)&1.50883&1.50882&1.50883&0.000013& 1.50892&1.50874 &1.50883& 0.000177\\
Q1(4.5)& 1.51138& 1.51136& 1.51137&0.000022&1.51153&1.51121&1.51137& 0.000317 \\ \hline
\multicolumn{9}{c}{4--2}\\ \hline
Q1(3.5)& 1.58694& 1.58692& 1.58693& 0.000014&1.58702&1.58693&  1.58684&0.000181\\
Q1(4.5) & 1.58974& 1.58972&  1.58973& 0.000025&  1.58989&1.58957& 1.58973& 0.000326\\ \hline
\multicolumn{9}{c}{5--3}\\ \hline
Q1(3.5) & 1.67397& 1.67324&1.67367&  0.00073&1.67334&1.67316& 1.67325& 0.000185 \\
Q1(4.5) & 1.67637& 1.67634& 1.67636&   0.000029 & 1.67652 &1.67619 & 1.67636& 0.000333 \\ \hline
\end{tabular}
\end{table*}

\subsubsection{Interline component}
\label{sec:ilc}

\begin{figure*}
\centering \includegraphics[scale=1.2]{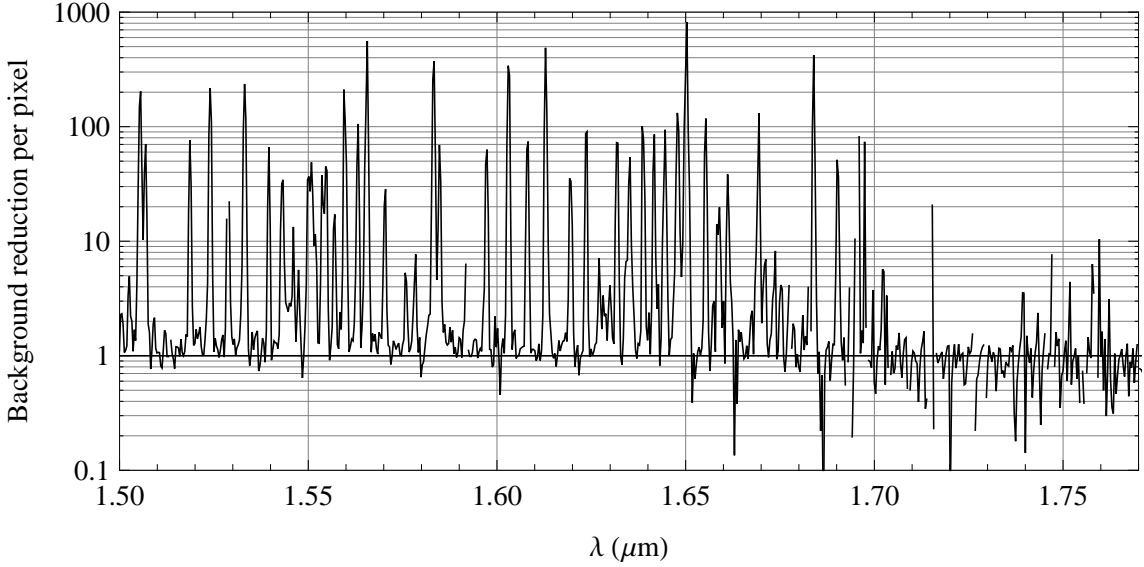}
\caption{The background suppression per pixel as a function of wavelength.}
\label{fig:backred}
\end{figure*}

We have measured the level of suppression at each pixel by dividing the control spectrum by the suppressed spectrum (after first correcting the control spectrum for the fact that this is from a single fibre compared to six fibres for the suppressed spectrum), this is shown in Figure~\ref{fig:backred}.
The mean reduction per pixel between $1.5$ and $1.7$ \um\ is $\approx 17$, and the median is $\approx 1.6$.  The reduction of the integrated background between $1.5$ and $1.7$ \um\ is $\approx 9$.

\begin{figure*}
\centering
\subfigure[OH suppressed]{
\centering \includegraphics[scale=0.6]{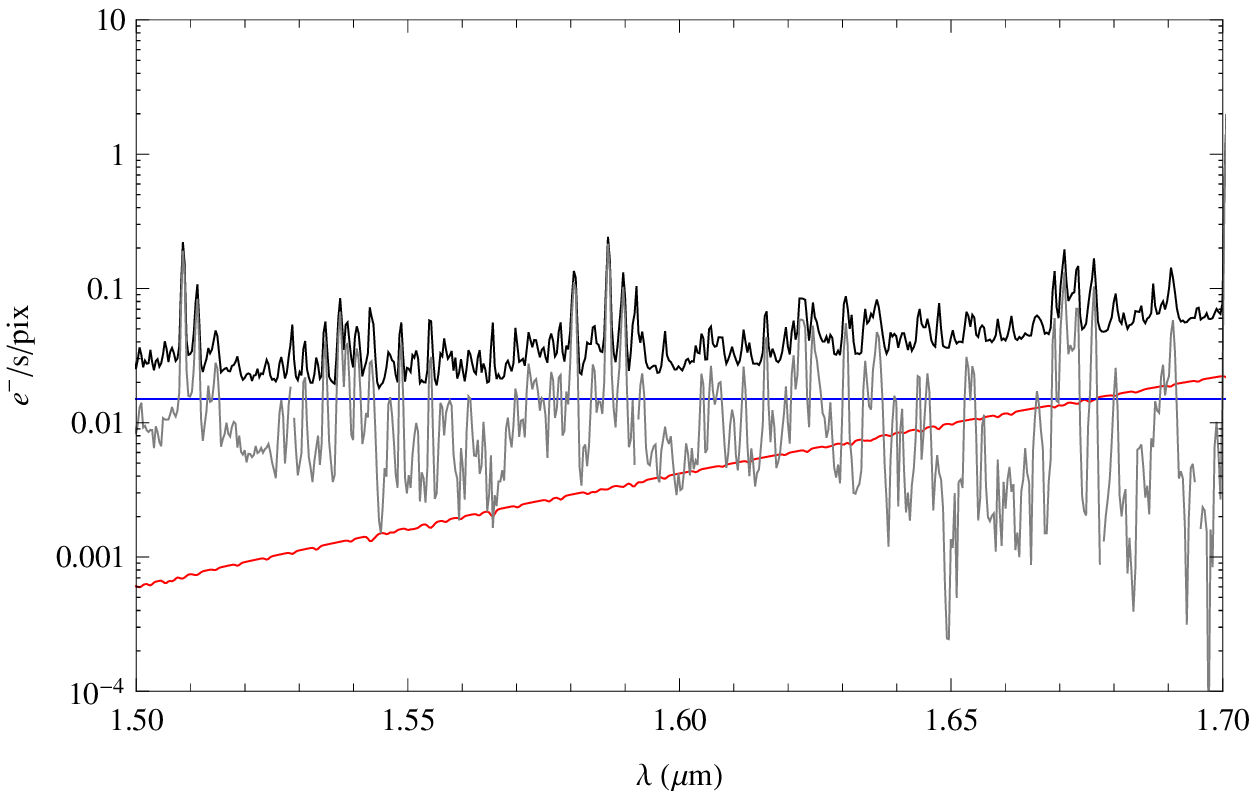}
}
\subfigure[Unsuppressed]{
\centering \includegraphics[scale=0.6]{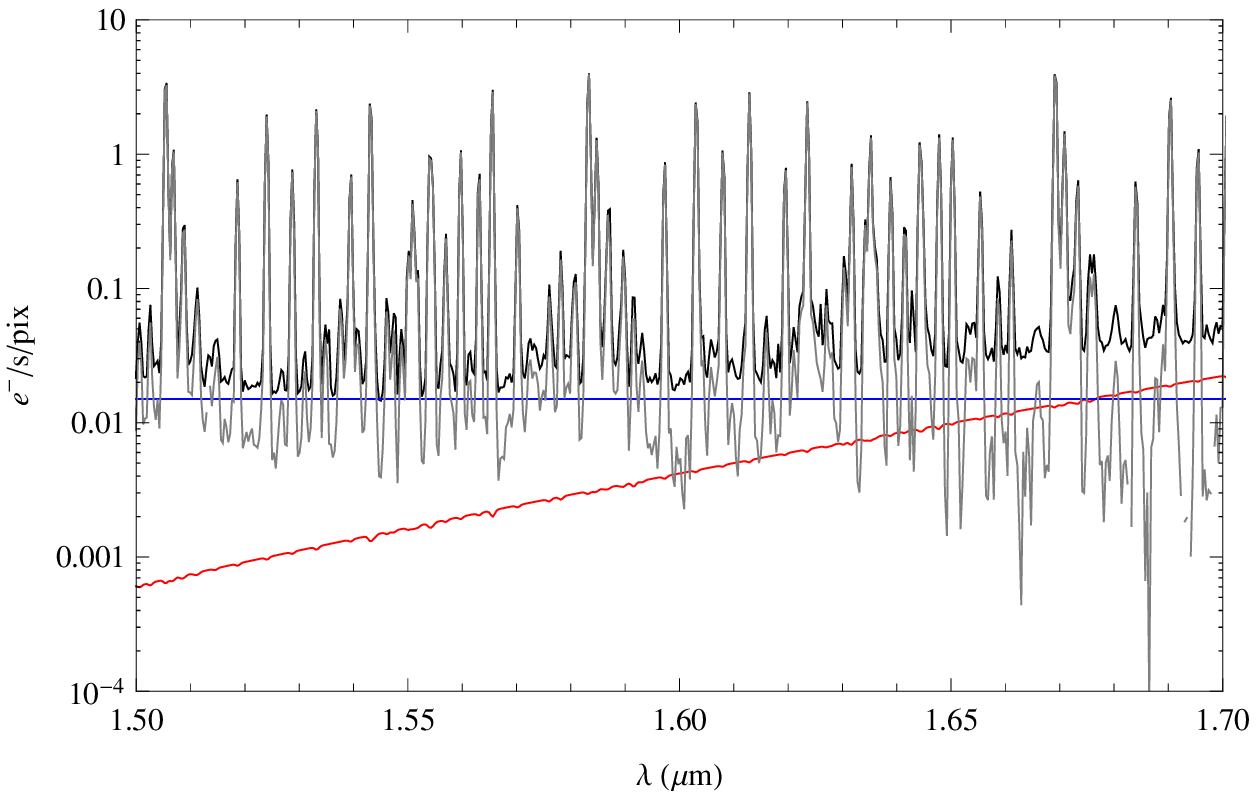}
}
\caption{The electrons per pixel per second from the dark current (blue), thermal emission (red) and night sky (grey) compared to the unsubtracted spectra (black) for suppressed and unsuppressed observations.  The dark current and thermal emission are significantly brighter than the interline component we are trying to measure.}
\label{fig:backmodel}
\end{figure*}

\begin{figure}
\centering \includegraphics[scale=0.6]{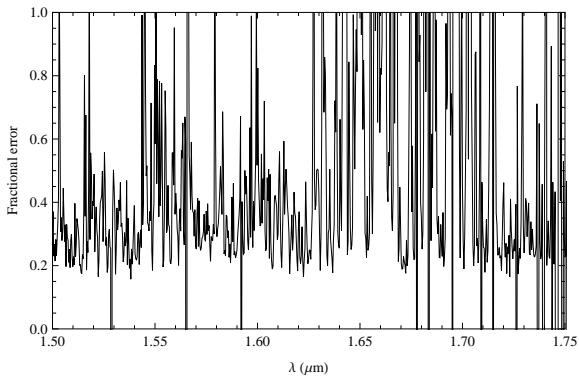}
\caption{The fractional error on the counts per pixel for the suppressed sky spectra, calculated as standard deviation / mean, for a half an hour exposure. }
\label{fig:error}
\end{figure}

To compare the interline components in the suppressed and control spectra we first subtract the detector dark current and instrument thermal emission.  However, these components are brighter than the interline component we are trying to measure, and the subtraction is inevitably noisy. 
 Figure~\ref{fig:backmodel} shows a comparison of the electrons per second per pixel due to dark current, thermal emission and suppressed and unsuppressed OH lines.  The statistical fractional error on the suppressed observations is shown in Figure~\ref{fig:error}, which is calculated from the mean and standard deviation of counts per pixel from15 separate half-hour exposures.  The typical error on the interline component is $\approx 25$ per cent; however there may be unquantified systematic errors on these measurements.

Nevertheless, after dark and thermal subtraction,
there is no discernible reduction in the background
between the OH lines.  
This is unexpected; the models of \citet{ell08} show that if the background (after dark subtraction) is composed of thermal emission, zodiacal scattered light and OH lines then the interline region should be dominated by light that originates as OH lines that is scattered by the spectrograph.  Thus 
suppressing the OH lines should also suppress this
light scattered from the OH lines.  The fact that we do not see this suggests four possibilities:  (i)  there is residual instrumental emission which has not been properly removed in the data reduction, (ii) the IRIS2 spectrograph does not scatter significant amounts of OH light between the lines, (iii) the interline region is dominated by some source that \citet{ell08} did not account for, (iv) the FBGs are not suppressing the OH lines as predicted, and hence the reduction is not as predicted. 

We have addressed (i), the possibility of residual thermal emission, as fully as possible in our data reduction method (\S~\ref{sec:datred}).  However we are operating in the regime of very low count rates, such that systematic errors may become significant.  The raw ADU per pixel in the interline component region is $\sim 10$ in a half hour exposure, which corresponds to ~$\sim 45$ e$^{-}$ per pixel, of which $\sim 18$ e$^{-}$ per pixel are due to the sky and the detector dark current contributes $\sim 27$ e$^{-}$ per pixel .  That is we are detecting less that 1 e$^{-}$ per pixel per minute, and we are dominated by detector noise.  At this level systematics become very important.  For example the effective read noise is $\approx 8$ e$^{-}$ per pixel  which alone gives $\sim 15$ per cent uncertainty in the interline component measurements.  There are systematics in detector linearity such as discussed in section~\ref{sec:cube}, and there may be other effects such as reciprocity failure (\citealt{bie11}) which are very difficult to characterise.

We can discount (ii) since we have measured the scattering of IRIS2, and indeed it was this empirical scattering model from IRIS2 that was used in the models of \citet{ell08}.  There is a small change in the PSF of IRIS2 due to a different slit and a different f-ratio at the input, however this will affect only the Gaussian core of the PSF, and then only slightly, and not at all the Lorentzian wings of the PSF.

\begin{figure}
\centering \includegraphics[scale=.6]{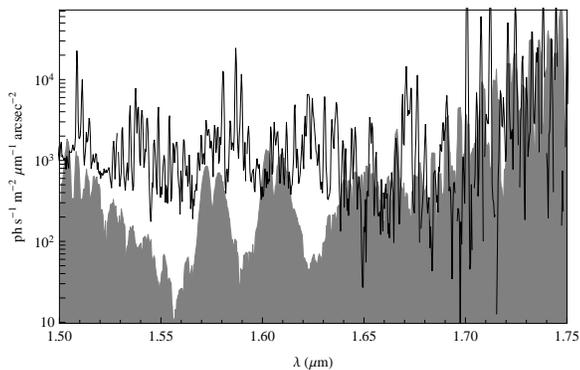}
\caption{The estimated molecular emission from the HITRAN2008 database (grey) compared to the GNOSIS interline component (black line).}
\label{fig:hitran}
\end{figure}

Regarding (iii), we have searched for other molecular bands and lines of emission in the H band.  We used the SpectralCalc.com facility to compute the emission line intensities from the HITRAN2008 models (\citealt{rot09}).  We included lines from H$_{2}$O, CO$_{2}$, O$_{3}$, N$_{2}$O, CO, CH$_{4}$, O$_{2}$, NO, SO$_{2}$, NO$_{2}$, NH$_{3}$, HNO$_{3}$, N$_{2}$, HF, HCl, HBr, HI, ClO, OCS, H$_{2}$CO, HOCl, HCN, CH$_{3}$Cl, H${_2}$O$_{2}$, C$_{2}$H$_{2}$, C$_{2}$H$_{6}$, PH$_{3}$, COF$_{2}$, SF$_{6}$, H$_{2}$S, HCOOH, HO$_{2}$ and ClONO$_{2}$.  The relative intensities were taken directly from the atmospheric paths of SpectralCalc.com assuming an observing altitude of 1.2 km, a target height of 600km, standard atmospheric conditions and zenith angle of 0 deg.  The resulting emission is shown in Figure~\ref{fig:hitran} where we have normalised the intensity to match roughly the interline component.  There are hints that some of the structure in the interline component arises from certain molecular bands.  The O$_{2}$ band at 1.58 $\mu$m and a CH$_{4}$ feature at $\approx 1.667\ \mu$m are most obvious, but there are possibly some other broad features due to CO$_{2}$, N$_{2}$O and C$_{2}$H$_{2}$.  However the relative intensities of these features are not right, and neither is the overall shape of the combined emission from these molecules.  Moreover, with the exception of O$_{2}$ and  CH$_{4}$ there are no features corresponding to the obvious emission lines seen throughout.  

\begin{figure}
\centering \includegraphics[scale=.6]{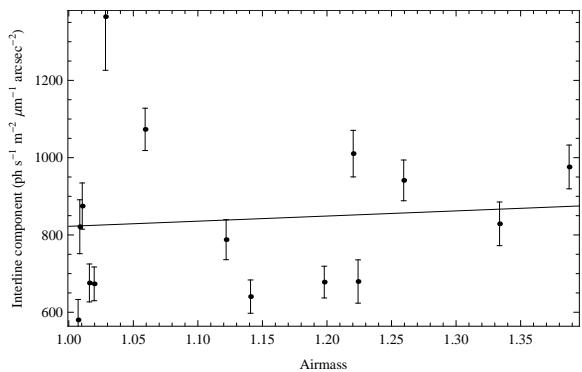}
\caption{The interline component strength as a function of airmass and the best fitting linear relation.  The error bars show the error on the mean interline component emission.}\label{fig:airmass}
\end{figure}

Scattering from aerosols or dust in the upper atmosphere, and certain chemical reactions may also produce a nightglow continuum.  For example the continuum produced in the reaction NO $+$ O $\to$ NO$_{2} +$ h$\nu$, has been found to correlate well with the optical nightglow (\citealt{ster72a,ster72}), and it has been suggested (\citealt{con96})that this may increase in the NIR and is of the same order as the brightness of the interline component measured by \citet{mai93}. We have measured the interline component strength for individual observations as a function of airmass and find no trend, see Figure~\ref{fig:airmass}.  This suggests that the interline component does not have an atmospheric origin, but we caution that the atmospheric emission could be temporally variable, as for the OH lines, which could mask any dependence on airmass.   In summary there are  individual features of the interline component due to other atmospheric molecular emission, but as yet we cannot explain the overall structure of the interline emission.

\begin{figure}
\centering \includegraphics[scale=.6]{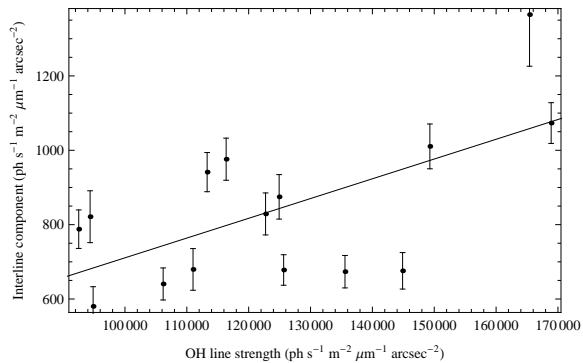}
\caption{The interline component strength as a function of OH line strength and the best fitting linear relation.  The error bars show the error on the mean interline component emission.}
\label{fig:ilcoh}
\end{figure}

Finally we consider (iv), that there may still be unsuppressed OH emission dominating the interline component (e.g.\ the Q1(3.5) and Q1(4.5) transitions discussed in \S~\ref{sec:ohsupp}).   We have tested this and find a weak correlation between the OH line strength and the interline component strength for individual observations (recall that the OH line strength varies by about 10 per cent on the timescale of minutes \citealt{ram92,fre00} and by a factor of $\approx 2$ throughout the night \citealt{shi70}), shown in Figure~\ref{fig:ilcoh}.  The r$^{2}$ goodness of fit is 0.38, with a significance of $p<0.02$.  Thus the OH line intensity does indeed affect the interline component.  \emph{Nota bene}, this is expected even with OH suppression, and is not an indication that the FBGs are not correctly suppressing the OH lines.  \cite{ell08} show that at a resolving power of $R=2000$ the scattering from residual suppressed OH lines, even though very faint, will still dominate the interline region even after OH suppression.  In fact it is the relative weakness of the correlation between interline component and OH line strength that must be explained, especially when coupled with the lack of dependence on airmass in Figure~\ref{fig:airmass}.  This will at least in part be due to uncertainty in the measurements.  The signal to noise on the interline component is poor, and the standard deviation of the residuals to the best fit shown in Figure~\ref{fig:ilcoh} is 160 \bright, which gives an indication as to the error on the interline component measurements.  However, the weakness of the correlation  may also be further indication that there is still some unidentified source of interline emission.

In summary, we feel confident that we can rule out (ii), i.e.\ we believe we understand the scattering function, but cannot rule out any of the other options.  Instrumental artifacts, unknown atmospheric components and inaccurate modelling of the OH lines may all contribute to the interline component at some level.

\subsection{Sensitivity}

\begin{figure}
\centering \includegraphics[scale=0.6]{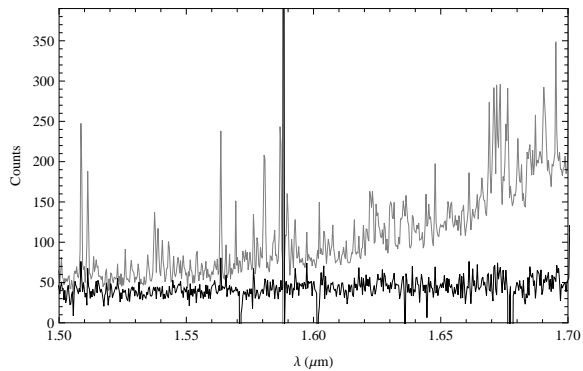}
\caption{The spectrum of the low surface brightness galaxy HIZOA J0836-43 from a half hour exposure (black), and the same spectrum including the background (grey).}
\label{fig:lsbg}
\end{figure}

We have measured the sensitivity of GNOSIS with an observation of a low surface brightness galaxy, which obviates the need to address the problem of aperture losses.  The galaxy selected was HIZOA J0836-43 (see \citealt{clu10}), with an H band surface brightness of 17.3 mag arcsec$^{-2}$, the details of which were kindly provided by Michelle Cluver.  We obtained half an hour exposure on target, the details of which are given in Table~\ref{tab:obs}.  The resulting spectrum is shown in Figure~\ref{fig:lsbg} along with a spectrum with no sky or dark subtraction.

\begin{figure}
\centering \includegraphics[scale=0.6]{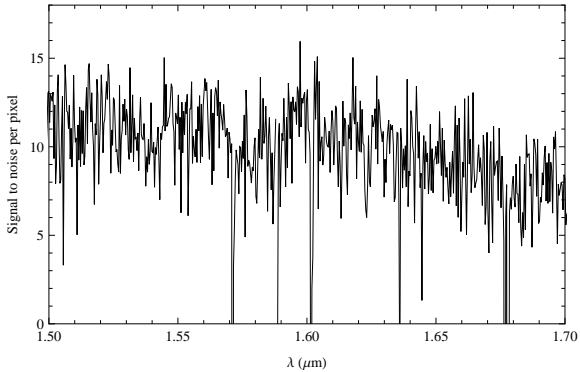}
\caption{The signal-to-noise per pixel for the observation of HIZOA J0836-43 shown in Figure~\ref{fig:lsbg}.}
\label{fig:snpix}
\end{figure}

\begin{figure}
\begin{center}
\subfigure[]{
\includegraphics[scale=0.6]{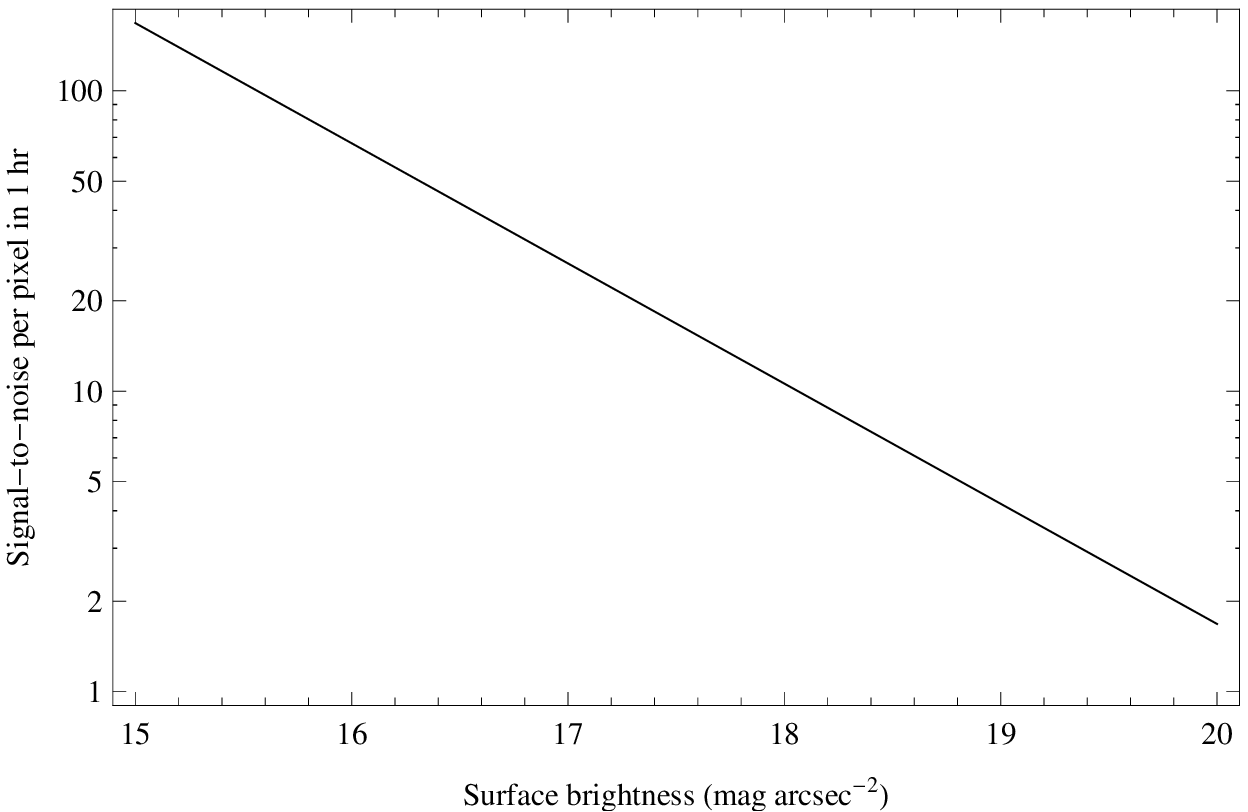}
}
\subfigure[]{
\includegraphics[scale=0.6]{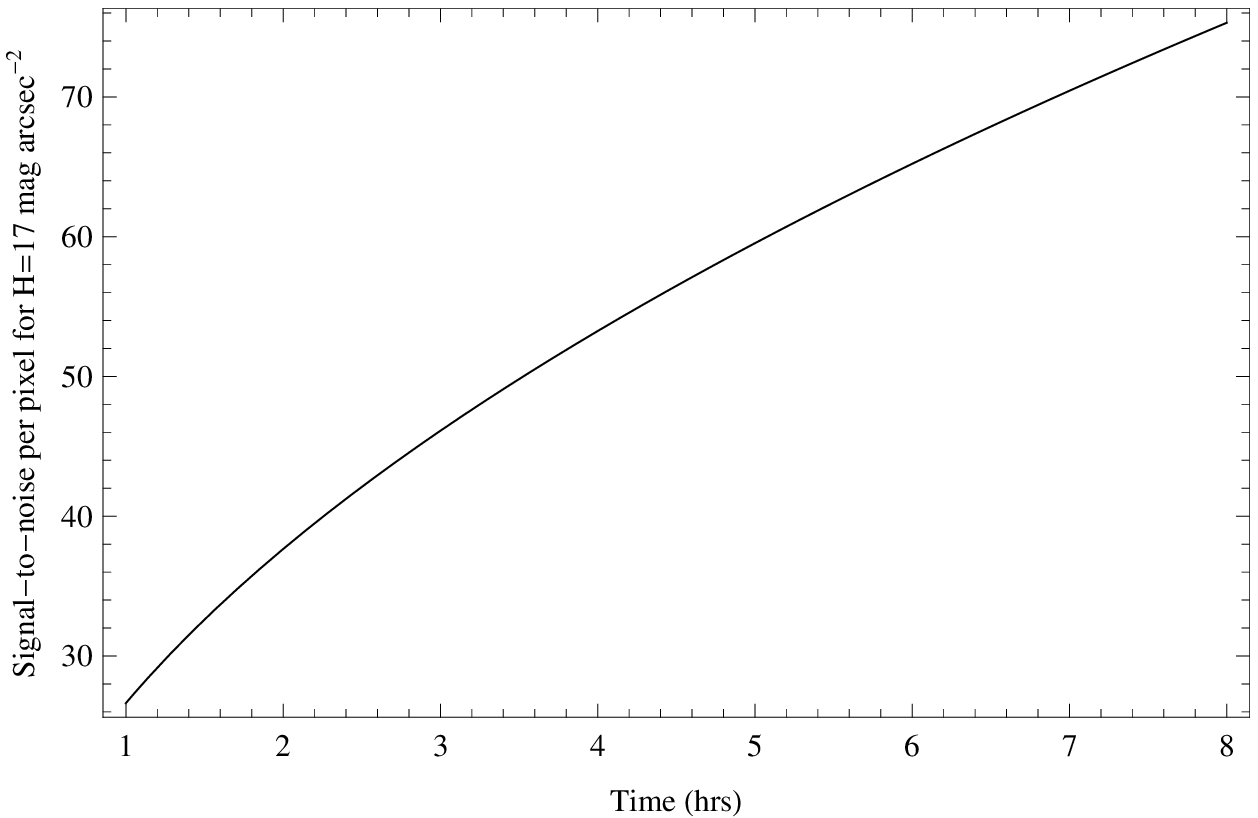}
}
\caption{The median signal-to-noise per pixel as a function of time and magnitude.}
\label{fig:sn}
\end{center}
\end{figure}

The signal-to-noise per pixel for this observation was calculated by dividing the sky-subtracted spectrum by the square root of the non-subtracted spectrum after first scaling by the detector gain and correcting to the full GNOSIS IFU area (as this spectrum is from only 6 of the 7 IFU elements).  The signal-to-noise per pixel is shown in Figure~\ref{fig:snpix}.  The median signal-to-noise per pixel is 10.1.  An identical analysis on the control fibre yields an identical median signal-to-noise.  This is because the OH suppression only improves the signal-to-noise near the night sky lines; between the lines the background is the same and these interline pixels dominate the calculation of the average signal-to-noise.  The improvement that comes from the OH suppression is counteracted by the loss in throughput from the FBG unit.  We note that this is not an inherent problem for OH suppression; improvements in throughput of the FBG unit along with lower thermal and detector backgrounds will improve the signal-to-noise.  A better understanding of the    interline component discussed in section~\ref{sec:ilc} may also lead to improvements in the suppression of the interline background as proposed in \citet{ell08}.  We have computed the median signal-to-noise as a function of time and magnitude, which are shown in Figure~\ref{fig:sn}.

\subsection{Illustrative observations}
\label{sec:sciobs}

We now demonstrate the benefit of OH suppression with three illustrative observations.  The first two are of [FeII] emission from the Seyfert galaxies, and allow a comparison of OH suppressed and non-OH suppressed observations. The third is of a candidate L7 dwarf star which is an example of OH suppressed observation of a continuum source.  The purpose of these examples is not to show that GNOSIS is competitive with the current generation of NIR spectrographs.  The prototype nature of GNOSIS results in a low overall throughput, a large number of pixels per square arcsecond and resultant high detector noise, and an efficiency loss in having to spend half the exposure time on blank sky, and therefore GNOSIS observations do not give any advantage over standard IRIS2 observations.  Rather, we wish to show the potential of OH suppression, by means of a comparison of the OH suppressed and control spectra.  With improvements to the throughput, the thermal background and the detector dark current similar observations of both emission line and continuum sources will be possible on much fainter objects.
We discuss how future instruments can best benefit from the potential of OH suppression and what changes are necessary to do so (\S~\ref{sec:disc}).

\begin{figure}
\begin{center}
\subfigure[NGC 7714]{
\includegraphics[scale=0.6]{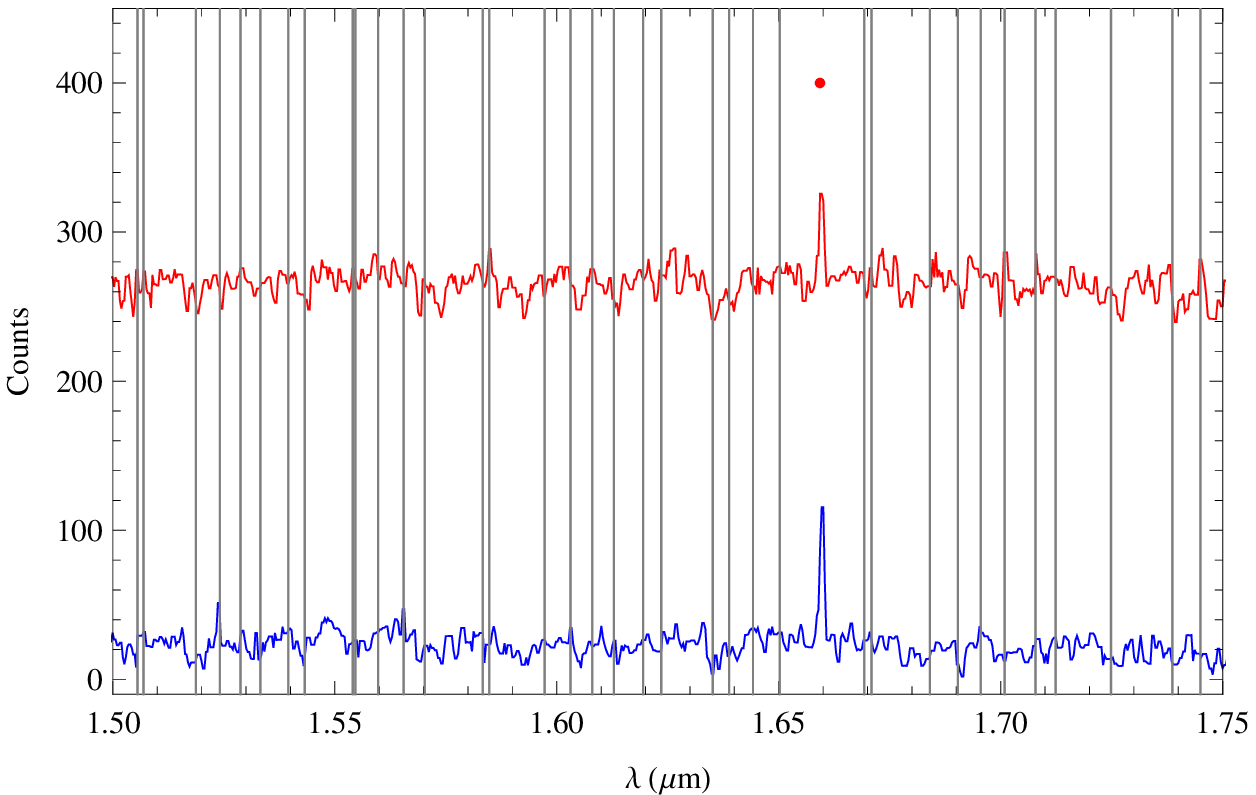}
}
\subfigure[NGC 7674]{
\includegraphics[scale=0.6]{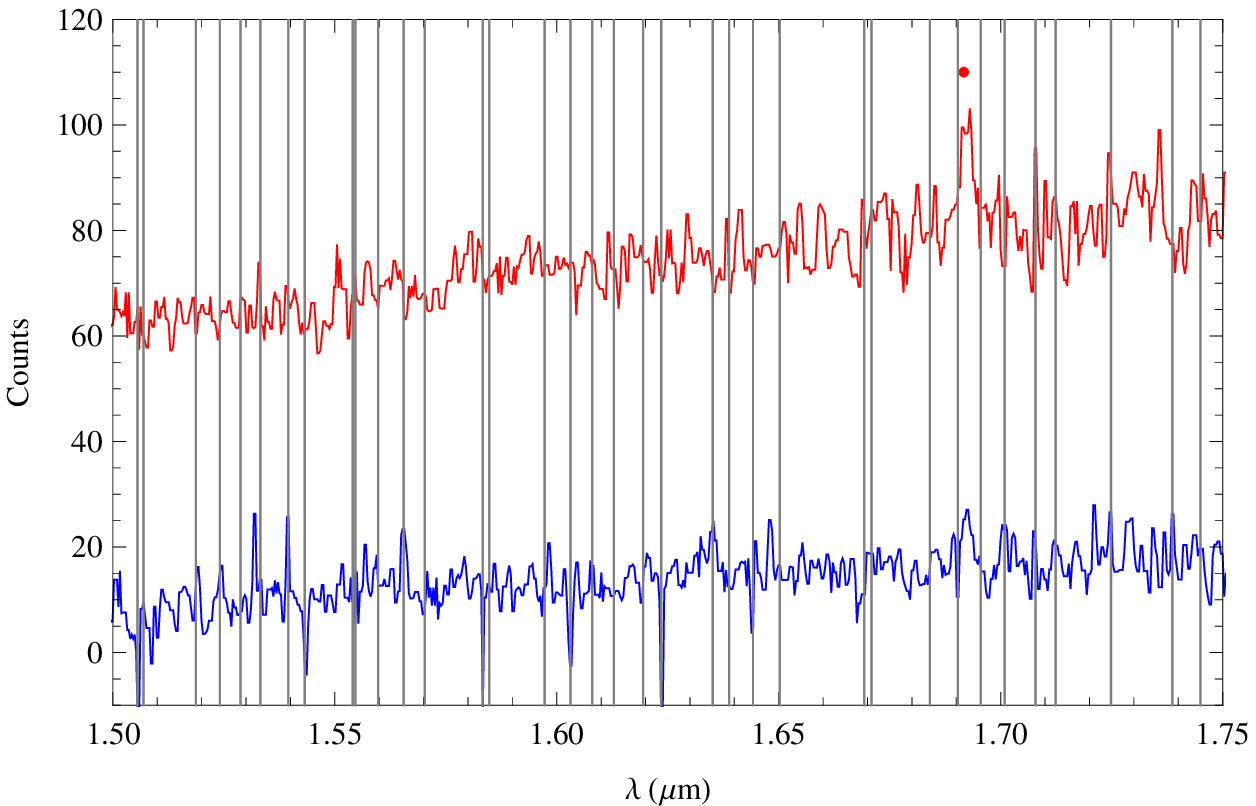}
}
\caption{The [FeII] emission from two Seyfert galaxies.  The black line is the OH suppressed spectrum from the middle fibre and the blue is the control fibre.  They have been offset for clarity.  The red dot marks the expected position of the 1.644 $\mu$m [FeII] line.  The grey  lines mark the positions of the brightest OH lines.}
\label{fig:seyferts}
\end{center}
\end{figure}

We give two examples of [FeII] emission from the Seyfert galaxies NGC~7674 and NGC~7714.  In both cases we compare a single OH suppressed fibre to the single control fibre, shown in Figure~\ref{fig:seyferts}.  In the case of NGC~7714 the [FeII] emission falls between the OH lines and is clearly visible in both suppressed and non-suppressed spectra.  In the case of NGC~7674 the [FeII] emission falls in a dense region of OH lines, but is still clearly visible in the suppressed spectrum, but less so in the non-suppressed spectrum.

\begin{figure}
\centering
\includegraphics[scale=0.6]{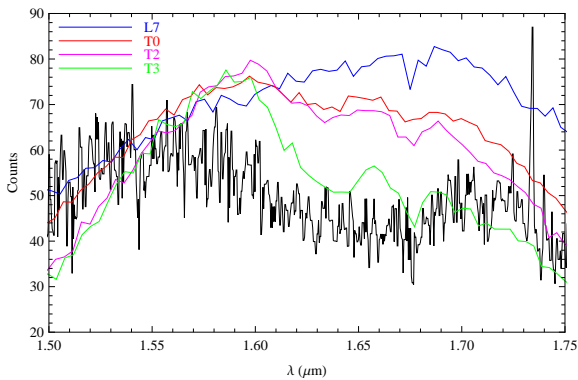}
\caption{A GNOSIS spectrum of an L7 dwarf from \citet{kir08} compared to standard star spectra from \citet{burg04}.  The H band spectrum shows stronger CH$_{4}$ absorption than expected for an L dwarf.}
\label{fig:l7}
\end{figure}

Figure~\ref{fig:l7} shows a GNOSIS spectrum of a candidate L7 dwarf, 2MASS J0257-3105, which was observed as part of Prof. C. Tinney's scheduled GNOSIS observing on 27th Nov 2011 (see Table~\ref{tab:obs}).  For these observations no control fibre was used, and therefore no comparison of OH suppressed spectra to non-suppressed can be made.  However, our experience observing brown dwarfs with IRIS2 tells us that the spectrum is comparable in quality and signal-to-noise to standard long slit IRIS2 spectroscopy.  The observations provide an example of the kind of science which may be done with OH suppression on continuum sources.  The object was  selected as an L7 dwarf from \citet{kir08}.  However, a comparison of the H band spectrum with standard spectra from \citet{burg04} shows that it has  stronger CH$_{4}$ absorption, similar to an early T dwarf.  An exact spectral type is difficult to give since none of the standard spectra gives a perfect match.

\section{Discussion}
\label{sec:disc}

We have demonstrated the first instrument to employ OH suppression with fibre Bragg gratings.  To test the performance of this new technology we made five key measurements: (i) the overall, and component, instrument throughput, (ii) the instrument sensitivity, (iii) the level of OH suppression, (iv) the interline component, (iv) illustrative observations of Seyfert galaxies.  We now discuss the results of these measurements in the context of the continuing development of OH suppressed NIR spectrographs.

(i) The total throughput is $\approx 4$ per cent but the throughput of IRIS2 is only $\approx 12$ per cent while that of GNOSIS itself is $\approx 36$ per cent.  The latter can be greatly improved by the following measures.
First on a specialised OH suppression spectrograph the need for relay optics could be eliminated by using a fibre vacuum feed through.  This would remove at least 4 optical surfaces, greatly improve alignment losses and have the additional benefit of lowering the thermal background.  Secondly, an even larger improvement could be made in the photonic lanterns.  Recent laboratory tests imply that the efficiency of the photonic lanterns is a function of the input f ratio, and that by feeding the lanterns at a slower speed we could increase the efficiency from 50 per cent to $\approx 70$ per cent.  These tests are preliminary, but indicate that the coupling efficiency of the lanterns may depend on the mode being coupled.

(ii)  The sensitivity of GNOSIS+IRIS2 is about that of IRIS2 when used in its standard spectrograph mode with a slit..  That is, in the particular implementation used for GNOSIS, the benefits of OH suppression are offset by lower throughput, higher detector background, higher thermal background and lower observing efficiency.  The low throughput has been discussed above and in section~\ref{sec:through}.  

The high detector background comes from the fact that a small field of view of 1.2 arcsec is being divided into 7 elements, and the spectra from each of these is being sampled by $\approx 2$ pixels, thus when extracting the spectra for each spectral pixel there is dark current and read noise from $\approx 14$ spatial pixels, compared to $\approx 2$ spatial pixels for the same sky area using IRIS2 (recall that GNOSIS feeds IRIS2 at a slower f-ratio than for long-slit IRIS2 observations).  Since the background is dominated by detector dark current this is a significant loss of sensitivity.  

The high thermal background results almost entirely from the slit block, and is a significant source of noise at $\lambda \gsim 1.65 \mu$m.
GNOSIS observations are also less efficient than IRIS2 observations in that half the time must be spent on sky, whereas with IRIS2 the object can be nodded up and down the slit.  Note that these problems are not intrinsic to OH suppression in general, and it is easy to envisage improved instruments that circumvent all these issues.

Using photonic lanterns with a larger number of cores will allow the use of larger diameter fibres and thus increase the field of view per fibre, thereby lessening the effects of detector background.  This will require faster spectrograph optics of around $\approx f/2.8$. 

The low observing efficiency is simply a matter of cost.  If two IFUs can be afforded then cross beam switching of observations will ensure that all observing time is spent on target.

Of course there are also many improvements that should be made to the spectrograph itself to improve the overall efficiency.  For example newer detectors with lower dark current and higher quantum efficiency, VPH gratings instead of grisms, and fixed format optics will all improve the throughput.  These improvements would of course be true of non-OH suppressed spectrographs as well.

(iii)  The suppression of the OH lines was a success.  The sky spectra and the object spectra show that the OH lines can be cleanly removed whilst maintaining good throughput (in the FBGs themselves) between the lines.  The level of suppression is high with $\approx 78$ per cent of the targeted lines suppressed at the required level.  Of the lines which do not meet the required level of suppression 37 per cent are due to  residuals from the Q1(3.5) and Q1(4.5) OH lines (\S~\ref{sec:ohsupp}).  Similarly there is an unsuppressed O$_{2}$ line.

(iv)  The expected reduction of the interline component was not observed.    This seems to indicate either an unaccounted source of interline emission, inaccuracy of our OH line models or unaccounted-for systematic errors.  

Systematic errors are a real possibility since we are operating in a very low-count regime of less that 1 e$^{-}$ per pixel per minute, and we are dominated by detector noise.  We have observed (and corrected for) systematic deviations from detector linearity as discussed in section~\ref{sec:cube}, and have noted that there may still be other effects such as reciprocity failure (\citealt{bie11}) which are very difficult to characterise.

We have already noted that specific transitions in the OH energy levels were not properly suppressed, which seems to be due to a larger energy gap between the $\Lambda$ doublets than predicted in our models.   It is possible that other such errors may exist in fainter lines, or that other branches of lines are brighter than supposed, but we are unable to measure this with the current sensitivity.  Unsuppressed faint OH lines could masquerade as continuum due to the spectrograph scattering.   

On the other hand there could be emission from another source.  A thorough search of the emission lines from 33 different molecules in the HITRAN2008 database cannot account for the structure and the faint lines we observe in the interline component except for a couple of individual features from O$_{2}$ and CH$_{4}$.  Also there is no dependence of the interline component on airmass, as would be expected for an atmospheric source.   However, there always remains the possibility that there is  another overlooked source or that or relative intensities of the molecular emission needs improving.

To try to distinguish between these two possibilities we looked at the correlation between the interline component and the OH line strength and found that they were only weakly correlated.  If true, this suggests that there is indeed an unknown source of interline emission, since otherwise the correlation would be much stronger.  However, this interpretation should be treated with much caution due to the large measurement errors and possibly significant systematic errors on the interline measurements.

The absolute level of the interline emission is $860 \pm 210$ \bright\ which is slightly higher than that measured by \citet{mai93}, (590 \bright) and slightly less than that measured by \citet{cub00} (1200 \bright).  We again point out that there may be significant systematic error on our measurement due to the low count rates measured.  However, the similarity to previous measurements could indicate that we have indeed reached a floor in the interline component which is significantly higher than the zodiacal scattered light background (see \citealt{ell08}).

If the interline emission is the result of unsuppressed OH lines due to inaccuracies in our OH line models, this may be fixed with better models or with wider FBG notches to account such uncertainties.  
However, as things stand, the lack of a reduction in the interline component suggests that OH suppression will find its niche in low resolution ($R \lsim 3000$) observations which would normally be insufficient to `resolve out' the OH lines.  Indeed OH suppression could be used at low resolutions, $R \sim 500$, at which conventional observations would be impractical.

(v)  The suggested benefit of OH suppression at low resolution is supported by our observations of Seyfert galaxies (\S~\ref{sec:sciobs}).  We find that in regions where the OH line density is high, as for NGC~7674, the low resolution of IRIS2 makes it difficult to identify the [FeII] emission in the control spectrum, but that the line is obvious in the suppressed spectrum.  Similarly when working between the lines as for NGC~7714, then the benefits of OH suppression are much less apparent since in this case we are limited by the detector dark current in both cases.

We conclude by noting that GNOSIS is the first step toward an OH suppression optimised spectrograph, and not a fully fledged facility instrument.  The overall performance of the OH suppression unit itself is good, but can be substantially improved. However there remains uncertainty about the origin of the interline emission, which could higher than predicted either due to some unknown source of emission or because of unaccounted-for systematics in the detector.  The next step will be to design and build a spectrograph optimised for a FBG feed, which will avoid the disadvantages that result from retrofitting an OH suppression system to an existing spectrograph.  Furthermore, in doing so we can capitalise on the lessons learnt with GNOSIS in terms of detector background, optimal spectral resolution and design.  

We present simulations of such a system in Figure~\ref{fig:sims}.  This simulation is based on the method presented in \citet{ell08} with modifications to reflect our experience with GNOSIS.  These modifications include an updated OH line spectrum which is an amalgamation of the observations of \citet{abr94} with the models of \citet{rous00}, giving precedence  to \citet{abr94} where lines occur in both lists.  The thermal background modelling has also been updated to consider the emission and the numerical aperture of each component of the optical train separately.  We first use the models to simulate the GNOSIS spectrum and compare these to the observations.
We next model PRAXIS, a concept for an optimised OH suppression system.  Specifically we assume that we have a corrected FBG design to account for the residual OH lines, that the fibre slit is cooled via a vacuum feed-through, that we have improved spectrograph throughput through the use of VPH gratings, and lower intrinsic detector noise through the use of a Hawaii-2RG 1.7$\mu$m cut-off detector, and better spatial sampling with 2 pixels per PSF FWHM.

\section*{Acknowledgments}
We thank Pierre Rousselot and Chris Lidman for illuminating discussions on the OH spectrum.
We thank Michelle Cluver for kindly providing the details of HIZOA J0836-43 for observation.   We thank the referee for useful comments which have improved this paper.
GNOSIS was funded by an ARC LIEF grant LE100100164.  This research has benefitted from the SpeX Prism Spectral Libraries, maintained by Adam Burgasser at http://pono.ucsd.edu/$\sim$adam/browndwarfs/spexprism.
 
\bibliographystyle{scemnras}
\bibliography{ps}

\end{document}